\documentclass[12pt]{article}
\usepackage{graphicx}
\usepackage{color}
\usepackage[normalem]{ulem} 

\begin{document}
\begin{center}
{\bf Test of the Orbital-Based LI3 Index as a Predictor
of the Height of the $^3$MLCT $\rightarrow$ $^3$MC
Transition-State Barrier for [Ru(N$^\wedge$N)$_3$]$^{2+}$ 
Polypyridine Complexes in CH$_3$CN}
\end{center}
\normalsize

\vspace{0.5cm}

\begin{center}
Ala~Aldin M.\ H.\ M.\ Darghouth$^*$, Denis Magero, and Mark E.\ Casida
\end{center}

\vspace{0.5cm}





\begin{center}
{\bf Abstract}
\end{center}

Ruthenium(II) polypyridine compounds often have a relatively long-lived triplet
metal-ligand charge transfer ($^3$MLCT) state, making these complexes useful
as chromophores for photoactivated electron transfer in photomolecular 
devices (PMDs).  As different PMDs typically require different ligands
and as the luminescence lifetime of the $^3$MLCT is sensitive to the 
structure of the ligand, it is important to understand this state and
what types of photoprocesses can lead to its quenching.  Recent work
has increasingly emphasized that there are likely multiple competing
pathways involved which should be explored in order to fully comprehend
the $^3$MLCT state.  However the lowest barrier that needs to be crossed
to pass over to the nonluminescent triplet metal-centered ($^3$MC) state 
has been repeatedly found to be a {\em trans} dissociation of the 
complex, at least in the simpler cases studied.  This is the fourth in
a series of articles investigating the possibility of an orbital-based
luminescence index (LI3, because it was the most successful of three)
for predicting luminescence lifetimes.  In an earlier study of bidentate
(N$^\wedge$N) ligands, we showed that the gas-phase $^3$MLCT $\rightarrow$ 
$^3$MC mechanism proceeded via an initial charge transfer to a single
N$^\wedge$N ligand which moves symmetrically away from the central ruthenium
atom, followed by a bifurcation pathway to one of two $^3$MC enantiomers.  
The actual transition state barrier was quite small and independent, to 
within the limits of our calculations, to the choice of ligand studied.  
Here we investigate the same reaction in acetonitrile, CH$_3$CN, solution 
and find that the mechanism differs from that in the gas phase in that 
the reaction passes directly via a {\em trans} mechanism.  This has
implications for the interpretation of LI3 via the Bell-Evans-Polanyi
principle.

\vspace{0.5cm}
 \begin{center}
 \includegraphics[width=0.8\textwidth]{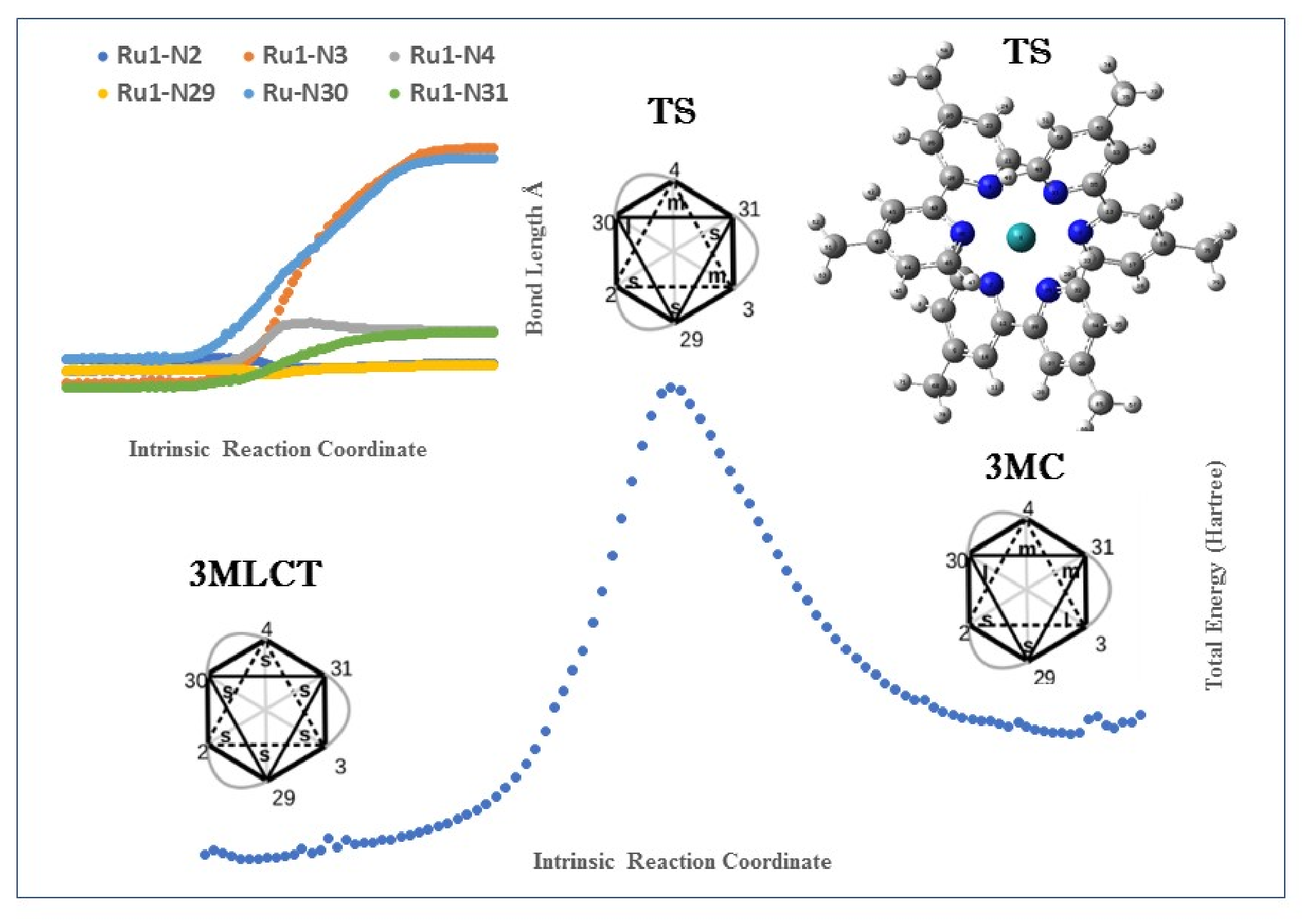}\\
 {\bf Graphical Abstract}
 \end{center} 

\vspace{0.5cm}

\section{Introduction}
\label{sec:intro}

\begin{figure}
\begin{center}
\includegraphics[width=0.9\textwidth]{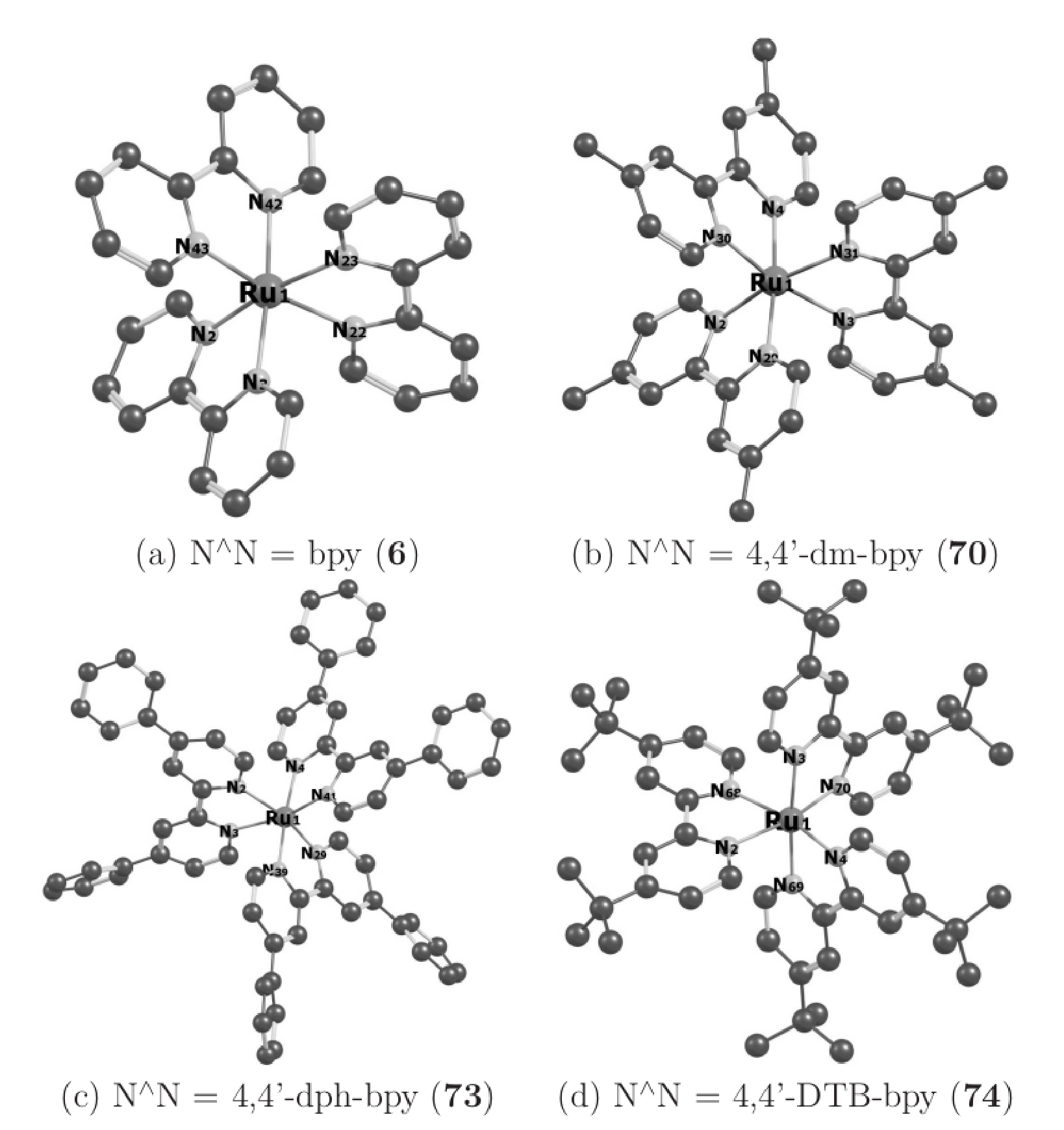}
\caption{Structures of [Ru(N$^\wedge$N)$_3$]$^{2+}$.  Hydrogen atoms
have been suppressed for clarity.  We have chosen to use
the $\Delta$ stereoisomers, though similar results are expected
for the corresponding $\Lambda$ stereoisomers.  Reproduced from 
Ref.~\cite{MDC24}.
\label{fig:complexstructures}
}
\end{center}
\end{figure}

\begin{figure}
\begin{center}
\includegraphics[width=0.9\textwidth]{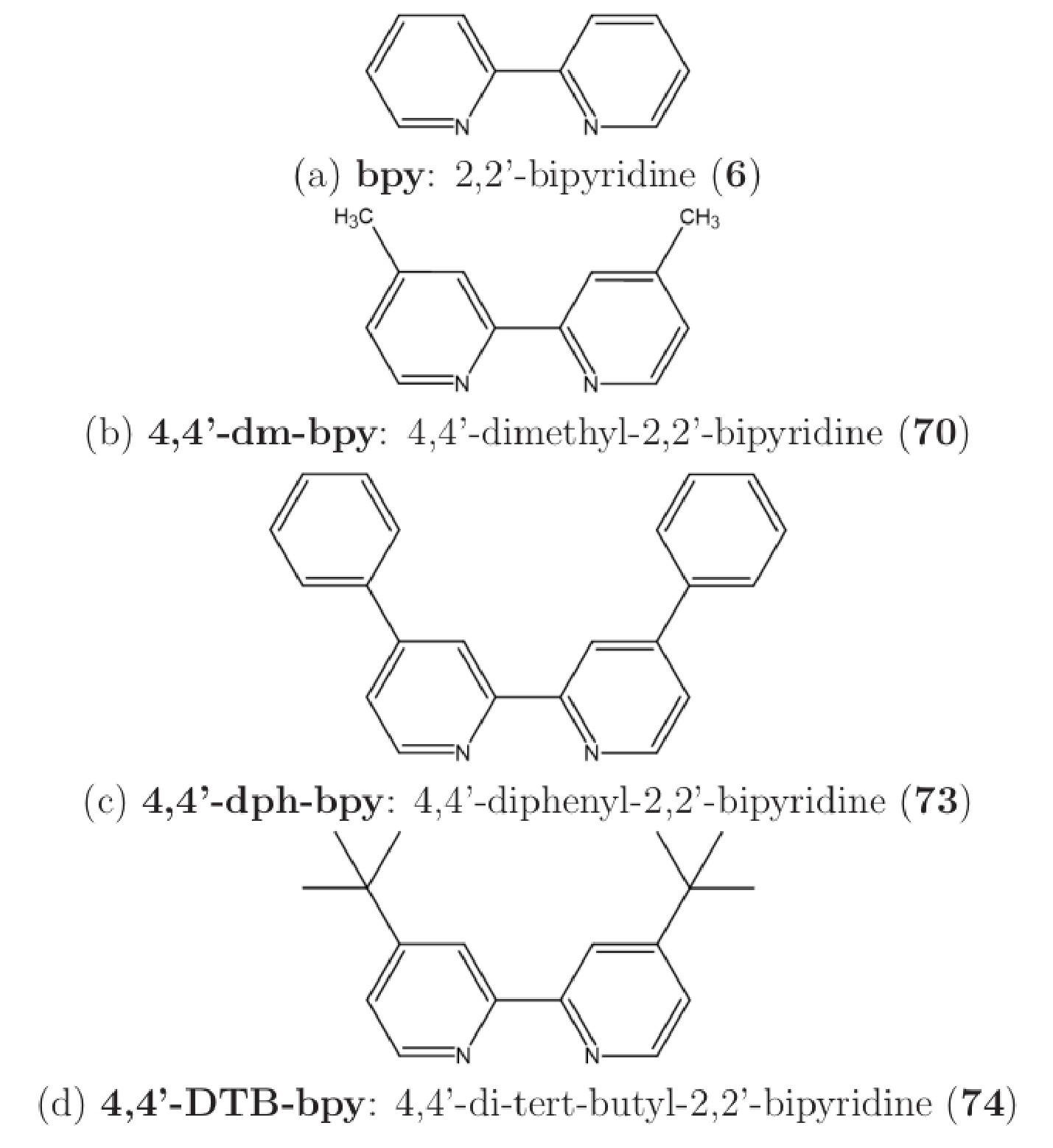}
\caption{Ligand list. Reproduced from Ref.~\cite{MDC24}.
\label{fig:ligandlist}
}
\end{center}
\end{figure}

Polypyridine ruthenium(II) complexes are a class of compounds containing
chromophores with a long-lived excited state which can transfer charge to
other molecules or within a single large molecule.  As such they have
elicited, and are expected to continue to elicit, immense interest
\cite{MMAC20,SCC+94,N82,LKW99,BJ01,DTL+03,HKZ03,MH05,%
NSF+08, JWW+10, SAH+20,BCCV21,HEG24}.  A central question has been what
features of the ligands favor long luminescence lifetimes and answers
are, at least informally, discussed within the context of ligand field theory
\marginpar{\color{blue} LFT}
(LFT).  This suggests that  we  should be able to understand luminescence
lifetimes not just from states but also, at least partially, from 
orbital information.  This orbital $\leftrightarrow$ luminescence lifetime
connection is gradually being elucidated in a series of articles, of which 
this is the fourth.  The first article (hereafter referred to as {\bf I}) 
\marginpar{\color{blue} PDOS, DFT}
showed how the partial density of states (PDOS)  obtained from 
gas-phase density-functional theory (DFT) calculations could be used to define
a LFT-like decomposition of the electronic structure of polypyridine 
ruthenium(II) complexes \cite{WJL+14}.  In particular, the energies 
$\epsilon$ of the ruthenium $t_{2g}$, $e_g^*$, and ligand $\pi^*$ orbitals 
could be determined with reasonable precision.  Article {\bf II} analyzed 
\marginpar{\color{blue} LI, LI3}
data for roughly one hundred compounds and found that an orbital-based 
luminescence index (LI) of which the third one,
\begin{equation}
  \mbox{LI3}= \frac{\left [ \left ( \epsilon_{{e}_{g}^{*}}
   +\epsilon_{\pi^{*}} \right )/2 \right ]^{2}}
   {\epsilon_{{e}_{g}^{*}}-\epsilon_{\pi^{*}}} \, .
  \label{eq:intro.1}
\end{equation}
correlated well with trends in experimental luminescence lifetimes as
\marginpar{\color{blue} $^3$MLCT, $^3$MC, $E_{\mbox{ave}}$}
represented by an empirical average triplet metal-ligand-charge-transfer 
($^3$MLCT) to triplet metal-centered ($^3$MC) energy barrier $E_{\mbox{ave}}$
 \cite{MCA+17}.
\marginpar{\color{blue} FMO}
The LI3 index was based upon frontier-molecular orbital (FMO) theory ideas
and designed with the $^3$MLCT $\rightarrow$ $^3$MC barrier height in mind
but this was not explicitly verified.  
Article {\bf III} carried 
out an investigation of the barrier to {\em trans} dissociation on the
lowest triplet potential energy surface (PES) for the compounds shown in
{\bf Fig.~\ref{fig:complexstructures}} for which line drawings are given
for the ligands in {\bf Fig.~\ref{fig:ligandlist}} \cite{MDC24}.  Calculations
in Article {\bf III} were gas-phase calculations using the same functional
and basis set as used in Article {\bf II}.  It was discovered that LI3
correlates very well with the $^3$MLCT-$^3$MC energy difference but that
\marginpar{\color{blue} TS}
finding the transition state (TS) for {\em trans} dissociation is complicated
by the presence of a bifurcation on the PES.  This bifurcation is, in part,
the result of the diversity of possible Jahn-Teller distorations in octahedral
complexes \cite{B01} which may give rise to similar competing product 
geometries \cite{BP23,C24}.  Once this bifurcation is taken
into account, barrier heights were found to be very similar for the four
compounds investigated.  However $E_{\mbox{ave}}$ was derived from condensed
phase data and the mechanism of charge transfer reactions can be sensitive
to the choice of solvent.  It is the objective of the present article to 
re-investigate LI3 and the {\em trans} dissociation mechanism on the triplet
PES using an implicit solvent model where the solvent has been chosen as
the polar solvent acetonitrile (CH$_3$CN, dielectric constant $\epsilon=37.5$),
very frequently used in experimental work on these complexes.  This means
\marginpar{\color{blue} $^1$GS, IRC}
that we must reoptimize the singlet ground state ($^1$GS), extract the PDOS,
recalculate LI3, reoptimize the two triplet minima ($^3$MLCT and $^3$MC),
and find the TS and intrinsic reaction coordinate (IRC) linking them for the
reaction in acetonitrile.

As this article is part of the special ``Computational Science
from Africa and the African Diaspora,'' it seems appropriate to
quote a proverb in the national language (Swahili) of the African author (DM):
\begin{quote}
\noindent
{\em Haraka haraka haina baraka}\\
Hurry hurry has no blessing.
\end{quote}
The relevance of this proverb in the present context lies in the complexity
of the problem of understanding luminescence lifetimes in ruthenium(II)
polypyridine complexes.  These complexes begin in a $d^6$ $^1$GS which
may be photoexcited to higher-lying states, notably singlet metal-ligand
charge-transfer ($^1$MLCT) states which is believed to decay rapidly via
radiationless relaxation to the lowest $^1$MLCT state is then quickly 
transformed to the phosphorescent $^3$MLCT state through intersystem
crossing due to the heavy atom effect. The luminescence of this
$^3$MLCT state is believed to be quenched by radiationless relaxation to
other states, notably by passing over a barrier to a $^3$MC state, thereby
leading to an {\em increase} in the rapidity of the disappearance of 
the luminescent state and hence to a {\em shorter} luminescence liftetime.  
This $^3$MC state is thought to further relax to the $^1$GS via another 
intersystem crossing.  In reality, there is probably not a single
$^3$MCLT $\rightarrow$ $^3$MC $\rightarrow$ $^1$GS pathway, but rather
many such pathways.  This is especially evident for heteroleptic complexes
as different ligands may undergo partial $^3$MCLT $\rightarrow$ $^3$MC
dissociation.  However even in homoleptic complexes, such as the ones
treated in this article, different types of $^3$MCLT $\rightarrow$ $^3$MC
dissociation are possible.  In their recent review article \cite{HEG24},
Hern\'andez, Eder, and Gonz\'alez summarize the literature by grouping
luminescence mechanism proposals into three types and then adding a fourth
type.  Type I says that the rate of luminescence quenching is determined 
soley by the $^3$MLCT-$^3$MC energy difference.  Type II says that the
rate of luminescence quenching is determined by the $^3$MLCT $\rightarrow$
$^3$MC barrier.  Type III takes into account the barriers for all three
of the processes $^3$MLCT $\rightarrow$ $^3$MC, $^3$MC $\rightarrow$ $^3$MLCT,
and $^3$MC $\rightarrow$ $^1$GS.  To this we may add a fourth type (type IV), 
as this is the subject of the Hern\'andez-Eder-Gonz\'alez article, which 
acknowledges that there are several different $^3$MC states and that
all of these states must be taken into consideration in order to obtain
a satisfactory theory.  However, the Hern\'andez-Eder-Gonz\'alez
theory is based upon Eyring's transition state theory which represents 
one limit of dynamical theory, Marcus theory being yet another limit,
and even the concept of temperature for photochemical reactions in excited
states is open to question.  In this sense, we find it interesting that
one theoretical study of [Ru(bpy)$_3$]$^{2+}$ in solution implied that 
charge is not transfered to a single ligand but rather to two ligands 
at a time \cite{TCR11} while another article suggests the importance 
of studying the volume of the various TSs and $^3$MC basins based upon 
a theoretical exploration of the photodynamics of rhodium(III) 
photosensitizers \cite{BP23}.
Nevertheless, we applaud the Hern\'andez-Eder-Gonz\'alez article and 
willingly acknowledge that the LI3 index proposed in Article {\bf II} is a 
type II theory which happens to fit trends derived from experiments on 
around 100 complexes fairly well.  But that our reasoning was indirect 
and not entirely satisfactory.  This led us to a further gas-phase
theoretical investigation (Article {\bf III}) which appears to favor LI3
as a type I theory.  However the experimentally-derived rates used
to develop LI3 were based upon condensed phase data.  This brings us
to the present article where we still consider only the
$^3$MC state resulting from {\em trans} partial dissociation according
to old rules given by Adamson for photodissociation of $O_h$ complexes
\cite{VC83}.  As ``hurry has no blessing,'' we slow down and make sure
that we understand what our orbital-based luminescence index, LI3, is
really telling us about the the Adamson mechanism in solution before
thinking about where we might go when all the complexities of a type IV
theory are taken into account.

The rest of this paper is organized as follows: Our computational methods
are described in the next section (Sec.~\ref{sec:details}).  
Section~\ref{sec:results} presents the results of our computational 
investigation.  Section~\ref{sec:conclude} contains our concluding
discussion.  Additional information is available as Supplementary Information 
\marginpar{\color{blue} SI}
(SI) (Sec.~\ref{sec:SI}).

%
%


\section{Methods}
\label{sec:details}

Our use of software was dictated not only by what was available to us,
but also, in part, by which author was most comfortable with which 
software.  It was thus important to verify that alll the software
gave (nearly) identical answers. AMHMD carried out calculations with
version G09 revision D.01 of {\sc Gaussian} \cite{g09} while DM carried
out calculations with version 5.0.4 of {\sc Orca} \cite{NWB+20}.  As
reproducibility is also a cornerstone of the scientific method, we also
wished to know just how were results obtained using the same method in
the two programs.  Of course, some differences are expected between different
programs using the same method just because of, for example, different 
convergence criteria, different grids, and different ways to construct 
solvent accessible volumes, but these are hopefully minor.

In Article {\bf III}, we addressed the problem of being able to carryout
the same calculation with {\sc Orca} as we had done in Articles {\bf I}
and {\bf II} with {\sc Gaussian}.  It turned out that specifying 
specifying the B3LYP \cite{SDCF94} functional does not mean
the same thing in the two programs.  When the VWN parameterization of
the local density approximation was programmed in {\sc Gaussian}, the
VWN parameterization of the random phase approximation (which we shall
\marginpar{\color{blue} VWN3, VWN5}
refer to as VWN3) was used even though this was not the orginal 
recommendation of Vosko, Wilke, and Nusair \cite{VWN80}.  Instead,
these authors recommended their parameterization (which we shall
refer to as VWN5) of the quantum Monte Carlo results of Ceperley
and Alder.  Up until {\sc Gaussian} used VWN for VWN3, VWN in the
literature generally meant VWN5.  Moreover the original B3LYP functional
programmed in {\sc Gaussian} uses the VWN3 parameterization, so we
\marginpar{\color{blue} B3LYP(VWN3)\\ B3LYP(VWN5)}
will call it B3LYP(VWN3) for clarity.  Later some programs redefined
B3LYP as B3LYP(VWN5).  For example, B3LYP in {\sc Orca} is B3LYP(VWN5)
but the keyword {\tt B3LYP/G} allows {\sc Orca} users to use B3LYP(VWN3).
We also used the {\sc Orca} keywords {\tt NORI} (no resolution-of-the-identity
approximation is used), {\sc TightSCF}, {\sc TightOpt}, {\sc SlowConv}, 
{\sc NumFreq} and an ultra-fine grid.
This was explicitly verified in Article {\bf III} where
\marginpar{\color{blue} NEB, IRC}
{\sc Orca} nudged elastic band (NEB) calculations were carried out using 
the B3LYP(VWN3) and the resultant NEB first estimate of the intrinsic 
reaction coordinate (IRC) was subsequently refined using {\sc Gaussian}
\cite{MDC24}.  

Articles {\bf I}-{\bf III} are only concerned about gas-phase calculations.
However the present article is specifically concerned with if and how our 
earlier conclusions obtained from gas-phase calculations will need to be
\marginpar{\color{blue} SMD}
changed within an implicit solvent model \cite{TP94,CT99,TMC05}.  
In particular, we are interested in the SMD (for Solvation Model based upon the 
quantum mechanical Density) for our implicit solvent model 
calculations \cite{MCT09} as this was used in the 
the NEB {\sc Orca} calculations carried out by the Toulouse 
group \cite{SDAH18} (See also Refs~\cite{LBS+12,DHAE17,SDAH18,SAH+18,SAH+20}).  

To this end, we chose one geometry of complex {\bf 6} ([Ru(bpy)$_3$]$^{2+}$) 
and tried to reproduce the same single point energies with {\sc Orca} 
and {\sc Gaussian} at a variety of computational levels.  
In particular, we explored the use of two different orbital basis sets and
\marginpar{\color{blue} ECP}
ruthenium effective core potentials (ECPs), namely the 6-31G basis set
\cite{DHP71,HDP72} combined with the LANL2DZ ECP on ruthenium \cite{HW85b}
and the larger def2-TZVP \cite{WA05} basis set combined with the 
Stuttgart SD28 ECP \cite{AHDP90,AHD+91}.  We also wished to include
Grimme's D3 dispersion correction \cite{GAEK10} with Becke-Johnson 
damping \cite{GEG11}.  

The keywords needed to do these calculations differ significantly between
{\sc Orca} and {\sc Gaussian}.  For example, setting 
{\tt EmpiricalDispersion=D3} in {\sc Gaussian} indicates to
use Grimme's original D3 dispersion correction \cite{GAEK10}.
The corresponding keyword in {\sc Orca} is {\tt D3ZERO}.
Setting {\tt EmpiricalDispersion=GD3BJ} in {\sc Gaussian} indicates
to use the Grimme's semiemprical D3 dispersion correction but with 
Becke-Johnson damping \cite{GEG11}.  The corresponding keyword in 
{\sc Orca} is {\tt D3BJ}.  
It is also possible for {\sc Gaussian} users to use B3LYP(VWN5) but the
procedure is more delicate and requires the following lines of input 
\cite{B3LYP5}.  
We also discovered that the {\tt SDD} keyword in {\sc Gaussian} gives
the Stuttgart SD28 ECP but uses a different orbital basis set, so that
the def2-TZVP for ruthenium needs to be made an explicit part of the input.
Thus, in order to do a B3LYP(VWN5)+D3BJ/def2-TZVP \&
Ru(SD28) calculation we used the following {\sc Gaussian} input:

\begin{verbatim}

#p  bv5lyp/gen nosymm  pseudo=read scrf=(solvent=acetonitrile,SMD) \\
gfinput pop=full iop(3/76=1000002000) iop(3/77=0720008000) \\
iop(3/78=0810010000) EmpiricalDispersion=GD3BJ

...

N C H
Def2TZVP
*********
Ru
Def2TZVP
*********

\end{verbatim}
Note that the backslashes here are just to indicate that the entry is
all on a single line and the ``...'' indicates the parts of the input
have been removed for brevety.
The results are shown in the SI.
The largest gas-phase error in the total energy is 67.5 $\mu$Ha (0.0424
kcal/mol) which is acceptable for the present work.

Of course our main interest in the present paper is if and how our
earlier gas-phase conclusions will need to be changed in the light of
SMD implicit solvent model calculations.  Both {\sc Gaussian} and {\sc Orca}
have implemented this method in a nearly identical way with gaussian
charges, rather than point charges.  However they differ
in the construction of the solvent accessible region.  In particular,
{\sc Gaussian} does an integral-equation-formalism (IEF) point charge continuum
\marginpar{\color{blue} IEF, PCM, C-PCM}
model (PCM) calculation \cite{CMT97,MCT97,MT97,TMC99} 
while {\sc Orca} does a conductor-like polarizable
continuum (C-PCM) \cite{KS93,TS95,BK97,BC98,CRSB03,GB03,GCD+08} calculation.  
A table in 
the SI indicates a difference
in total energies between {\sc Gaussian} and {\sc Orca} calculations of
as much as 600 $\mu$Ha (0.377 kcal/mol) which might be compared with the
maximum error of 0.5 kcal/mol (796 $\mu$Ha) for solvation energies
quoted for differences between the IEF-PCM and C-PCM implementations 
of the SMD quoted at the top of the second column on page~6390 of 
Ref.~\cite{MCT09}.  However a maximum error of less then 0.2 kcal/mol 
(318 $\mu$Ha) is claimed for ions in a solvent with dielectric constant
greater than 32 which is the case for acetonitrile (CH$_3$CN, $\epsilon 
\approx 35$).  This difference between the SMD calculations
in the two programs turns out to be controllable by varying the fineness
of the Lebedev grid used in {\sc Orca}. The precise keywords to use are
\begin{verbatim}
%cpcm
num_leb X
end
\end{verbatim}
where X is the number of Lebedev grid points per cavity sphere.  
The default is 110 in version 5 of {\sc Orca} but a table in the SI 
shows that the difference between {\sc Gaussian} and {\sc Orca} calculations 
may be considerably reduced by using a finer Lebedev grid.  We have 
therefore chosen to use a 590 point Lebedev grid in all our SMD 
calculations with {\sc Orca} unless otherwise specified. 
Of course, a comparison of energy differences calculated with
{\sc Gaussian} with the same energy differences calculated with {\sc Orca}
may even be smaller than our observed difference in total energies
as error cancellation is common when calculating energy differences.

A question that could be asked, but which is beyond the scope 
of this article, is to what extent the solvent model needs to
be modified to handle excited states which are typically more
diffuse and charged solutes, notably the $^3$MLCT state where
the positive charge is expected to be delocalized onto the ligands
rather localized on the ruthenium atom and shielded by the ligands
as in the $^3$MC state?  This, of course, is a question which is
general when using any implicit solvent model to treat excited
states in solution, with the case of a ``solvated'' electron,
such as when carrying out Birch reduction with sodium in liquid
ammonia, as a worst case scenario.  We are not aware of it having
been as yet much much discussed in the literature.

We carried out calculations at the 
following levels using the SMD implicit solvent 
model \cite{MCT09} and corresponding parameters for actonitrile (CH$_3$CN):
\begin{enumerate}
  \item B3LYP(VWN3)/6-31G \cite{DHP71,HDP72} \& LANL2DZ(Ru) \cite{HW85b}/SMD(CH$_3$CN):
        Ground state geometries, partial density of states (PDOS), luminescence index (LI),
        optimized $^3$MLCT and $^3$MC minima
        for all 4 complexes ({\bf 6}, {\bf 70}, {\bf 73}, and {\bf 74}) and NEB calculations
        for complexes {\bf 6}, {\bf 70}, and {\bf 73} as well as IRC calculations for complexes
        {\bf 6} and {\bf 70}.
  \item B3LYP(VWN3)+D3BJ/6-31G \& LANL2DZ(Ru)/SMD(CH$_3$CN):
        Ground state geometries, partial density of states (PDOS), luminescence index (LI),
        for all 4 complexes ({\bf 6}, {\bf 70}, {\bf 73}, and {\bf 74}).
  \item B3LYP(VWN3), 6-311G(d,p)=6-311G** \cite{KBSP80} \& LANL2DZ(Ru)/SMD(CH$_3$CN):
        Optimized $^3$MLCT and $^3$MC minima for complexes {\bf 73} and {\bf 74}.
  \item B3LYP(VWN3)+D3BJ/6-311G(d,p)=6-311G** \cite{KBSP80} \& 
        LANL2DZ(Ru)/SMD(CH$_3$CN)/SMD(CH$_3$CN): 
        Ground state geometries, partial density of states (PDOS), luminescence index (LI),
        for all 4 complexes ({\bf 6}, {\bf 70}, {\bf 73}, and {\bf 74}). 
\end{enumerate}


\section{Results and Discussion}
\label{sec:results}

\subsection{Ground State Geometries, PDOS, and LI3}

\begin{figure}
\begin{center}
\includegraphics[width=0.65\textwidth]{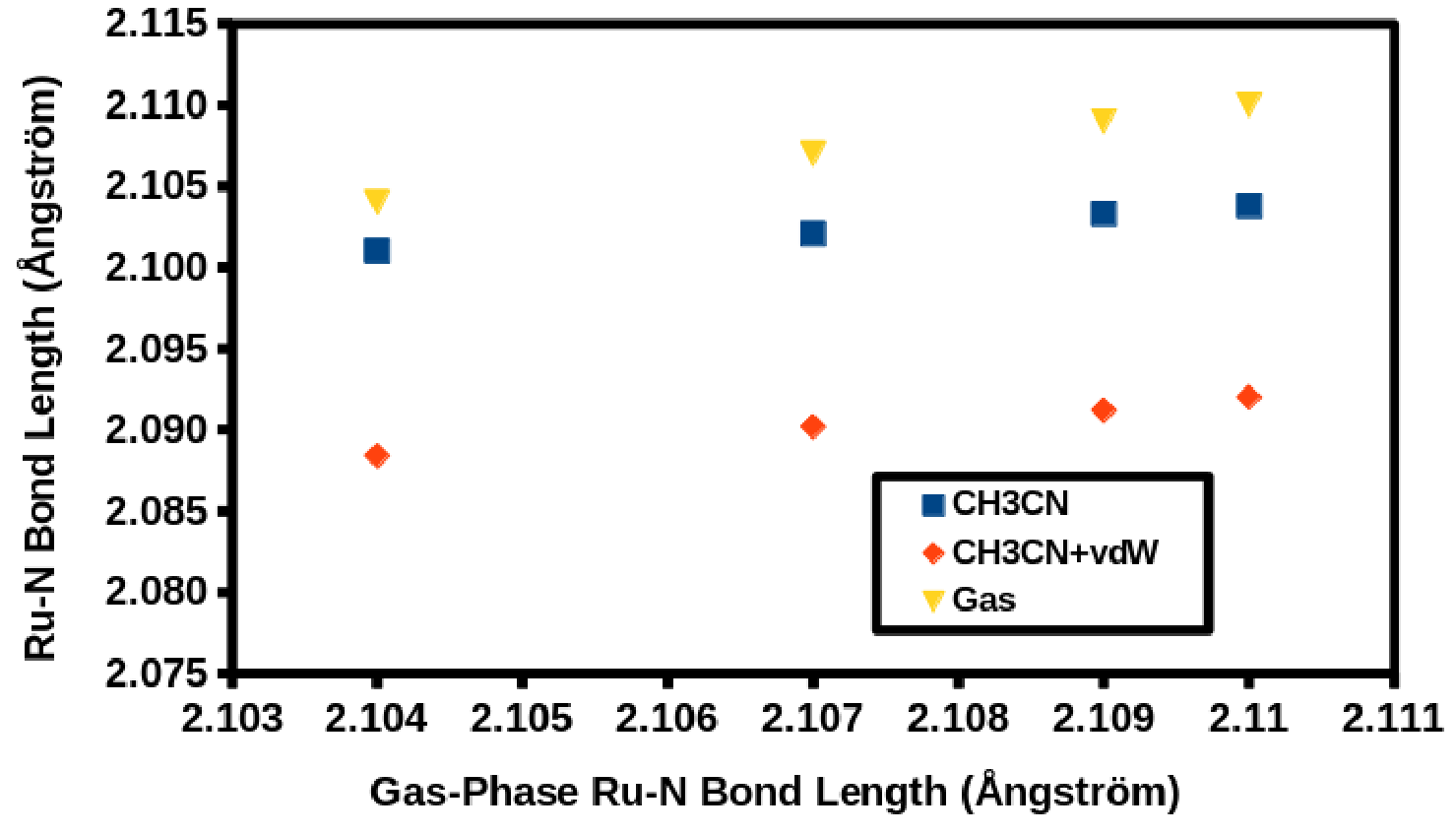}
\caption{Ground state average Ru-N bond lengths:
``Gas,'' B3LYP(VWN3)/6-31G \& LANL2DZ(Ru) (from \cite{MDC24}); 
``CH3CN,'' B3LYP(VWN3)/6-31G \& LANL2DZ(Ru)/SMD(CH$_3$CN);
``CH3CN+vdW,'' B3LYP(VWN3)+D3BJ/6-311G(d,p) \& LANL2DZ(Ru)/SMD(CH$_3$CN).
\label{fig:GSbondlength}
}
\end{center}
\end{figure}
Calculations at the B3LYP(VWN3)/6-31G \& LANL2DZ(Ru)/SMD(CH$_3$CN)
and \newline B3LYP(VWN3)+D3BJ/6-311G(d,p)\& LANL2DZ(Ru)/SMD(CH$_3$CN) 
levels are compared with gas-phase B3LYP(VWN3)/6-31G \& LANL2DZ(Ru) 
from Article {\bf III}.
As expected for a closed-shell symmetric molecule, the six ground state 
Ru-N bond lengths within each complex are essentially identical.
{\bf Figure~\ref{fig:GSbondlength}} shows that the complex
contracts slightly in going from the gas-phase to the condensed phase
(``chemical pressure'' \cite{LLY+22}).  An even larger compression results when
\marginpar{\color{blue} vdW}
the D3BJ (van der Waals, vdW) dispersion correction is included.


\begin{table}
\begin{center}
\begin{tabular}{cccc}
\hline \hline
\multicolumn{4}{c}{PDOS Energies} \\
HOMO & $t_{2g}$ & $1\pi^*$ & $e_g^*$ \\
\hline \hline
\multicolumn{4}{c}{Compound {\bf 6}} \\
\hline
\multicolumn{4}{c}{B3LYP(VWN3)/6-31G \& LANL2DZ(Ru) gas phase} \\
-11.20 eV & -11.06 eV & -7.42 eV & -5.01 eV \\
\multicolumn{4}{c}{B3LYP(VWN3)/6-31G \& LANL2DZ(Ru)/SMD(CH$_3$CN)}\\
-5.645 eV & -5.805 eV & -2.188 eV & +0.285 eV \\
\multicolumn{4}{c}{B3LYP(VWN3)+D3BJ/6-311G(d,p) \& LANL2DZ(Ru)/SMD(CH$_3$CN)}\\
-5.864 eV & -6.022 eV & -2.339 eV & +0.059 eV \\
\hline \hline
\multicolumn{4}{c}{Compound {\bf 70}} \\
\hline
\multicolumn{4}{c}{B3LYP(VWN3)/6-31G \& LANL2DZ(Ru) gas phase} \\
-10.42 eV & -10.55 eV & -7.00 eV & -4.53 eV \\
\multicolumn{4}{c}{B3LYP(VWN3)/6-31G \& LANL2DZ(Ru)/SMD(CH$_3$CN)}\\
-5.470 eV & -5.617 eV & -2.078 eV & +0.356 eV \\
\multicolumn{4}{c}{B3LYP(VWN3)+D3BJ/6-311G(d,p) \& LANL2DZ(Ru)/SMD(CH$_3$CN)}\\
-5.679 eV & -5.850 eV & -2.277 eV & +0.083 eV \\
\hline \hline
\multicolumn{4}{c}{Compound {\bf 73}} \\
\hline
\multicolumn{4}{c}{B3LYP(VWN3)/6-31G \& LANL2DZ(Ru) gas phase} \\
-9.84 eV & -9.88 eV & -6.53 eV & -4.04 eV \\
\multicolumn{4}{c}{B3LYP(VWN3)/6-31G \& LANL2DZ(Ru)/SMD(CH$_3$CN)}\\
-5.571 eV & -5.678 eV & -2.267 eV & +0.370 eV \\
\multicolumn{4}{c}{B3LYP(VWN3)+D3BJ/6-311G(d,p) \& LANL2DZ(Ru)/SMD(CH$_3$CN)}\\
-5.805 eV & -5.896 eV & -2.519 eV & +0.074 eV \\
\hline \hline
\multicolumn{4}{c}{Compound {\bf 74}} \\
\hline
\multicolumn{4}{c}{B3LYP(VWN3)/6-31G \& LANL2DZ(Ru) gas phase} \\
-10.08 eV & -10.20 eV & -6.61 eV & -4.19 eV \\
\multicolumn{4}{c}{B3LYP(VWN3)/6-31G \& LANL2DZ(Ru)/SMD(CH$_3$CN)}\\
-5.464 eV & -5.600 eV & -2.050 eV & +0.420 eV \\
\multicolumn{4}{c}{B3LYP(VWN3)+D3BJ/6-311G(d,p) \& LANL2DZ(Ru)/SMD(CH$_3$CN)}\\
-5.666 eV & -5.831 eV & -2.224 eV & +0.166 eV \\
\hline \hline
\end{tabular}
\end{center}
\caption{HOMO and PDOS orbital energies.
\label{tab:PDOSdata}
}

\end{table}



\begin{table}
\begin{center}
\begin{tabular}{ccc}
\hline \hline
Method & $\Delta_{\mbox{PDOS-LFT}}$ & Ru-N Bond Length \\
\hline 
\multicolumn{3}{c}{Compound {\bf 6}} \\
B3LYP(VWN3)/6-31G{\&}LANL2DZ(Ru) gas phase$^a$      & 48,800 cm$^{-1}$ & 2.110 {\AA} \\
B3LYP(VWN3)/6-31G{\&}LANL2DZ(Ru)/SMD(CH$_3$CN)                & 49,120 cm$^{-1}$ & 2.1038(4) {\AA} \\
B3LYP(VWN3)+D3BJ/6-311G(d,p){\&}LANL2DZ(Ru)/SMD(CH$_3$CN) & 49,050 cm$^{-1}$ & 2.0920(1) {\AA} \\
\hline 
\multicolumn{3}{c}{Compound {\bf 70}} \\
B3LYP(VWN3)/6-31G{\&}LANL2DZ(Ru) gas phase$^a$      & 48,600 cm$^{-1}$ & 2.109 {\AA} \\
B3LYP(VWN3)/6-31G{\&}LANL2DZ(Ru)/SMD(CH$_3$CN)           & 48,180 cm$^{-1}$ & 2.1033(4) {\AA} \\
B3LYP(VWN3)+D3BJ/6-311G(d,p){\&}LANL2DZ(Ru)/SMD(CH$_3$CN) & 47,850 cm$^{-1}$ & 2.0912(3) {\AA} \\
\hline 
\multicolumn{3}{c}{Compound {\bf 73}} \\
B3LYP(VWN3)/6-31G{\&}LANL2DZ(Ru) gas phase$^a$      & 47,100 cm$^{-1}$ & 2.104 {\AA} \\
B3LYP(VWN3)/6-31G{\&}LANL2DZ(Ru)/SMD(CH$_3$CN)           & 48,780 cm$^{-1}$ & 2.1010(6) {\AA} \\
B3LYP(VWN3)+D3BJ/6-311G(d,p){\&}LANL2DZ(Ru)/SMD(CH$_3$CN) & 48,150 cm$^{-1}$ & 2.0884(6) {\AA} \\
\hline 
\multicolumn{3}{c}{Compound {\bf 74}} \\
B3LYP(VWN3)/6-31G{\&}LANLD2Z(Ru) gas phase$^a$         & 48,500 cm$^{-1}$ & 2.107 {\AA} \\
B3LYP(VWN3)/6-31G{\&}LANLD2Z(Ru)/SMD(CH$_3$CN)             & 48,550 cm$^{-1}$ & 2.1021(13) {\AA} \\
B3LYP(VWN3)+D3BJ/6-311G(d,p){\&}LANL2DZ(Ru)/SMD(CH$_3$CN) & 48,370 cm$^{-1}$ & 2.0902(17) {\AA} \\
\hline \hline
\end{tabular}
\end{center}
$^a$From Ref.~\cite{MCA+17}.
\caption{PDOS-LFT splittings and average Ru-N bond lengths.
\label{tab:Delta}
}
\end{table}

\begin{figure}
\begin{center}
\includegraphics[width=0.9\textwidth]{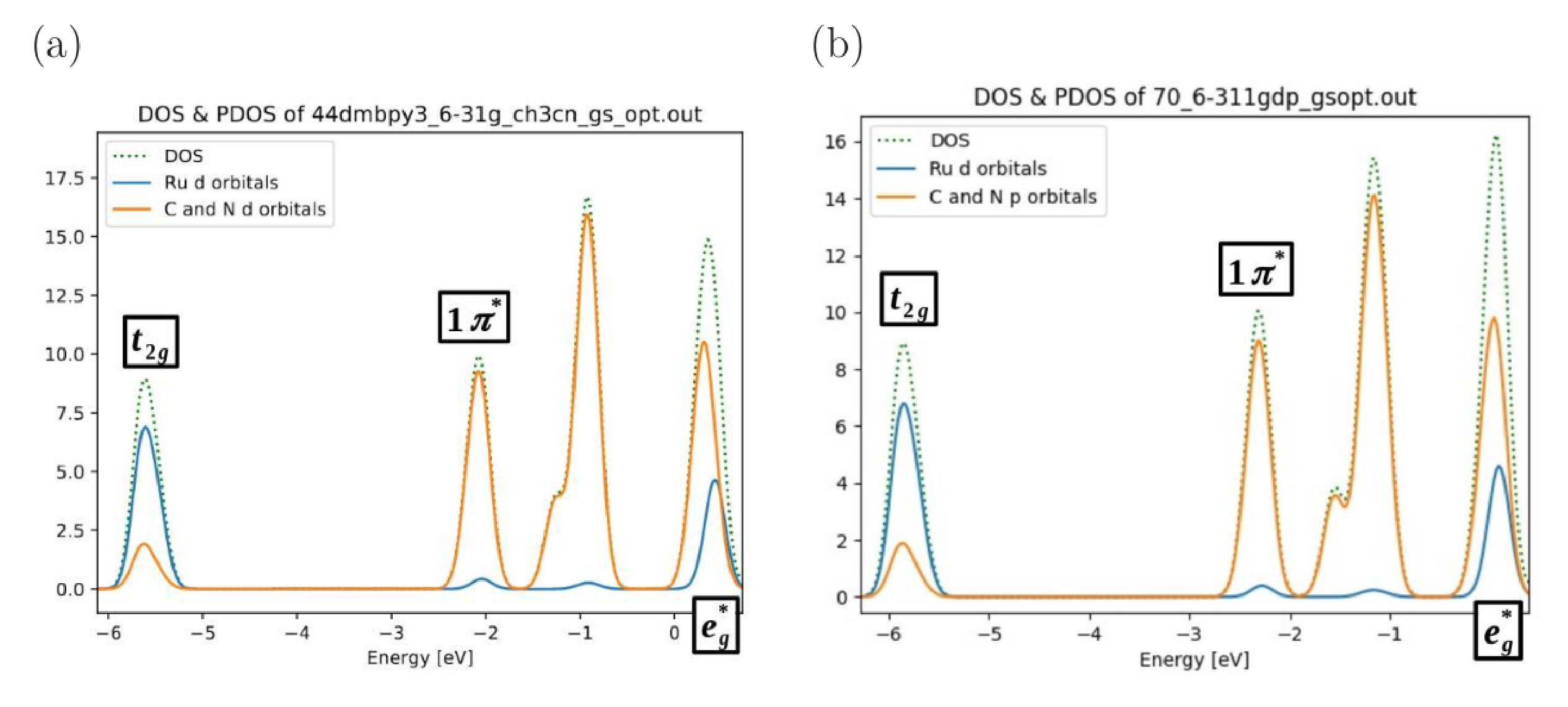}
\caption{
PDOS for complex {\bf 70} calculated with 40,000 points
and FWHM = 0.25 eV: 
(a) B3LYP(VWN3)/6-31G \& LANL2DZ(Ru)/SMD(CH$_3$CN),
(b) B3LYP(VWN3)+D3BJ/6-311G(d,p) \& LANL2DZ(Ru)/SMD(CH$_3$CN).
The corresponding gas-phase B3LYP(VWN3)/6-31G \& LANL2DZ(Ru) gas phase PDOS
may be found in Ref.~\cite{MCA+17}.
\label{fig:70_PDOS}
}
\end{center}
\end{figure}
\begin{figure}
\begin{center}
\includegraphics[width=0.90\textwidth]{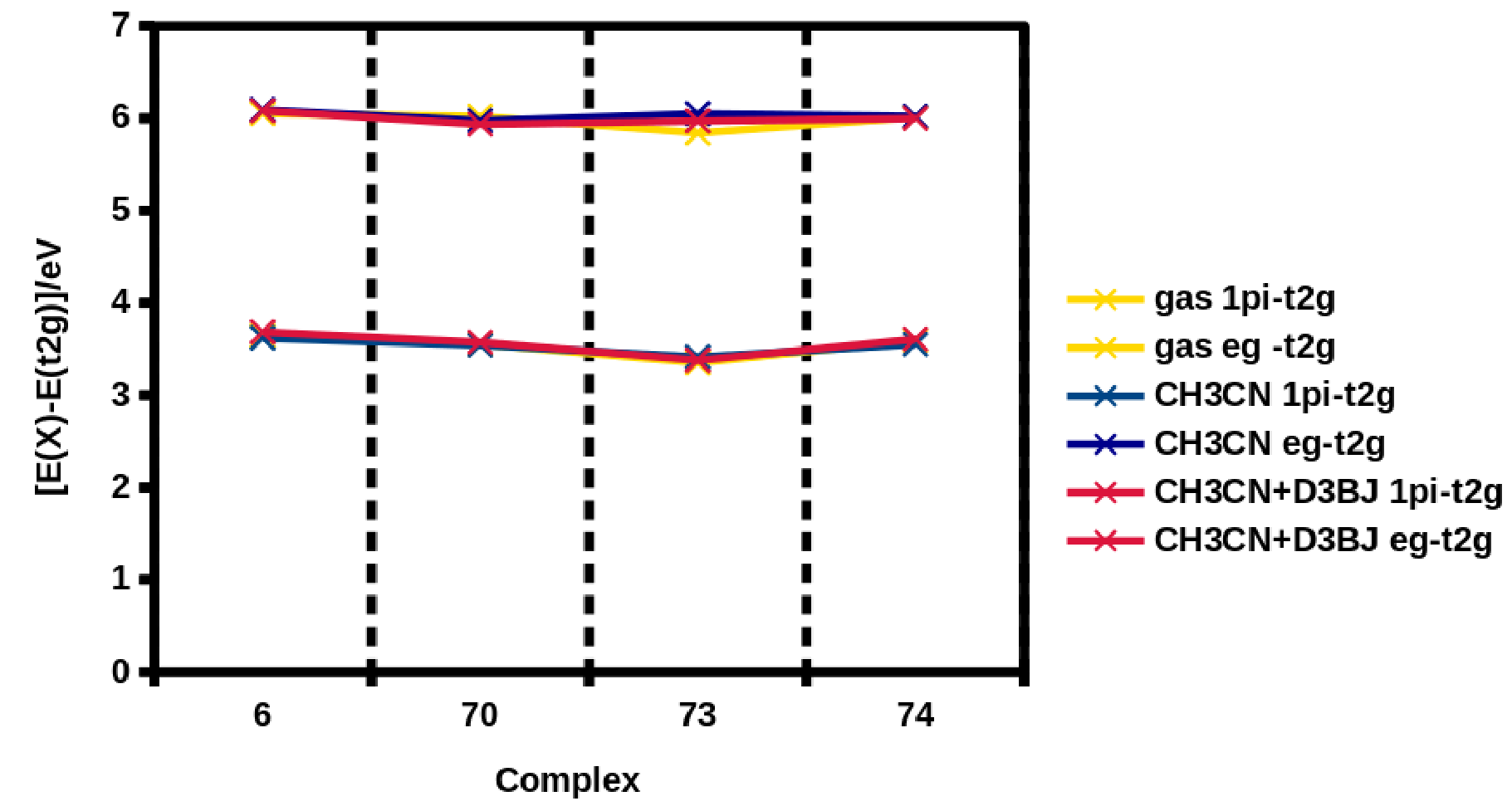} 
\caption{
Scaled PDOS orbital energies for all four complexes relative to the
$t_{2g}$ energy at each level of modeling:
``gas,'' B3LYP(VWN3)/6-31G \& LANL2DZ(Ru) (from \cite{MDC24}); 
``CH3CN,'' B3LYP(VWN3)/6-31G \& LANL2DZ(Ru)/SMD(CH$_3$CN);
``CH3CN+D3BJ,'' B3LYP(VWN3)+D3BJ/6-311G(d,p) \& LANL2DZ(Ru)/SMD(CH$_3$CN).
\label{fig:LFTenergies}
}
\end{center}
\end{figure}

The PDOS for complex {\bf 70} is shown in {\bf Fig.~\ref{fig:70_PDOS}}
while the PDOSs for the other complexes have been relegated to the SI.
{\bf Table~\ref{tab:PDOSdata}} give the HOMO and PDOS orbital energies
in the gas phase and with the SMD implicit solvation model.  Solvation
leads to a significant increase in orbital energies but there is remarkably
little change in orbital energy differences upon introduction
of solvent effects or solvent effects plus dispersion corrections as shown in 
{\bf Fig.~\ref{fig:LFTenergies}}.  {\bf Table~\ref{tab:Delta}}
shows that the order of the average Ru-N bond lengths is model independent
[R(Ru-N): {\bf 73} $<$ {\bf 74} $<$ {\bf 70} $<$ {\bf 6}].  It is therefore
normal to expect that the PDOS LFT splitting $\Delta_{\mbox{PDOS LFT}}$
should also vary in the order {\bf 73} $>$ {\bf 74} $>$ {\bf 70} $>$ {\bf 6}.
The fact that we do not really see this is most likely due to our inability
to calculate $\Delta_{\mbox{PDOS LFT}}$ with sufficient precision.


\begin{table}
\begin{center}
\begin{tabular}{cccc}
\hline \hline
\multicolumn{4}{c}{LI4} \\
{\bf 6} & {\bf 70} & {\bf 73} & {\bf 74} \\
\hline
\multicolumn{4}{c}{$E_{\mbox{ave}}$$^a$} \\
132 cm$^{-1}$ & 157 cm$^{-1}$ & 94 cm$^{-1}$ & 110 cm$^{-1}$ \\
\multicolumn{4}{c}{B3LYP(VWN3)/6-31G \& LANLDZ(Ru) gas phase$^b$} \\
16.03 eV   & 13.46 eV   & 11.21 eV   & 12.05 eV \\ 
\multicolumn{4}{c}{B3LYP(VWN3)/6-31G \& LANLDZ(Ru)/SMD(CH$_3$CN)}\\
17.12 eV & 13.87 eV & 10.32 eV & 11.94 eV \\
\multicolumn{4}{c}{B3LYP(VWN3)+D3BJ/6-311G(d,p) \& LANLDZ(Ru)/SMD(CH$_3$CN)}\\
17.49 eV & 13.93 eV & 10.66 eV & 12.40 eV \\
\hline \hline
\end{tabular}
\end{center}
$^a$Tables~10 and 11 of Ref.~\cite{MCA+17}.\\
$^b$From Ref.~\cite{MCA+17}.
\caption{Values of LI4 calculated with different functionals and basis
sets in gas phase and in CH$_3$CN.  
\label{tab:LI4}
}

\end{table}

\begin{figure}
\begin{center}
\includegraphics[width=0.90\textwidth]{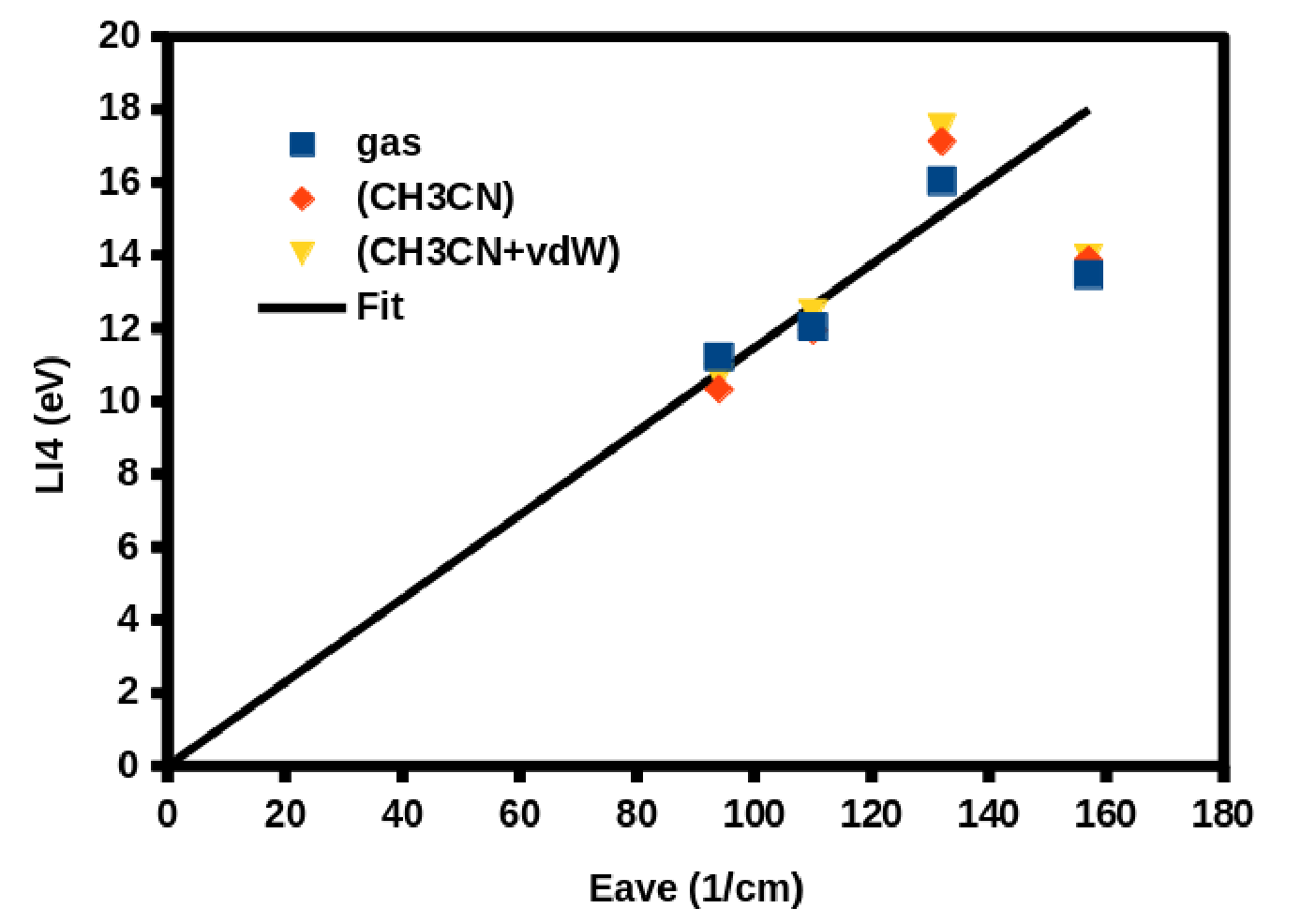} 
\caption{
Correlation of LI4 calculated using different levels of modeling
with $E_{ave}$ from Article {\bf II}:
``gas,'' B3LYP(VWN3)/6-31G \& LANL2DZ(Ru) (from \cite{MDC24}); 
``(CH3CN),'' B3LYP(VWN3)/6-31G \& LANL2DZ(Ru)/SMD(CH$_3$CN);
``(CH3CN+vdW),'' B3LYP(VWN3)+D3BJ/6-311G(d,p) \& LANL2DZ(Ru)/SMD(CH$_3$CN).
\label{fig:LI4}
}
\end{center}
\end{figure}

Let us turn to a modified orbital-based luminescence index LI4 shown in 
{\bf Table~\ref{tab:LI4}}.  Our observation that the PDOS in solution
is just shifted with respect to the PDOS in the gas phase suggests
that Eq.~(\ref{eq:intro.1}) should be modified to include an energy zero
$\epsilon_0$,
\begin{equation}
  \mbox{LI4}= \frac{\left ( \frac{\epsilon_{{e}_{g}^{*}}
   +\epsilon_{\pi^{*}}}{2} -\epsilon_0 \right )^{2}}
   {\epsilon_{{e}_{g}^{*}}-\epsilon_{\pi^{*}}} \, .
  \label{eq:results.1}
\end{equation}
Unfortunately different choices of $\epsilon_0$ will give different results.
This is because LI3 was motived by frontier molecular orbital
which depend upon the use of the Wolfsberg-Helmholz approximation \cite{WH52},
albeit with the additional assumption that the overlap integral is unity.
Since the Wolfsberg-Helmholz formula has a dependence on the energy zero, 
so does LI3.  Normally the Wolfsberg-Helmholz formula is used in extended 
H\"uckel theory where it is used to calculate matrix elements from gas-phase 
ionization potentials.  This might be an argument to explain
why LI3 constructed from gas-phase data seemed to work well for estimating 
trends in (condensed phase) 
luminescence lifetimes (Article {\bf II}).  We have tried different seemingly
logical choices of $\epsilon_0$ in Eq.~(\ref{eq:results.1}) and finally
concluded that the best choice is,
\begin{equation}
  \epsilon_0 = \epsilon_{\mbox{solvated}}^{\mbox{HOMO}} 
                - \epsilon_{\mbox{gas phase}}^{\mbox{HOMO}} \, ,
  \label{eq:results.2}
\end{equation}
which shifts the energies calculated in solvent to be close to those 
calculated in the gas phase.  This finishes the definition of LI4, 
as calculated and tabulated in Table~\ref{tab:LI4}.  
{\bf Figure~\ref{fig:LI4}} shows that this
definition gives very similar results at all levels of calculation
and correlated reasonably well with the quantity $E_{ave}$ derived
from experimental data in Article {\bf II}.

\subsection{$^3$MLCT and $^3$MC Minima}

\begin{figure}
\begin{center}
\includegraphics[width=0.90\textwidth]{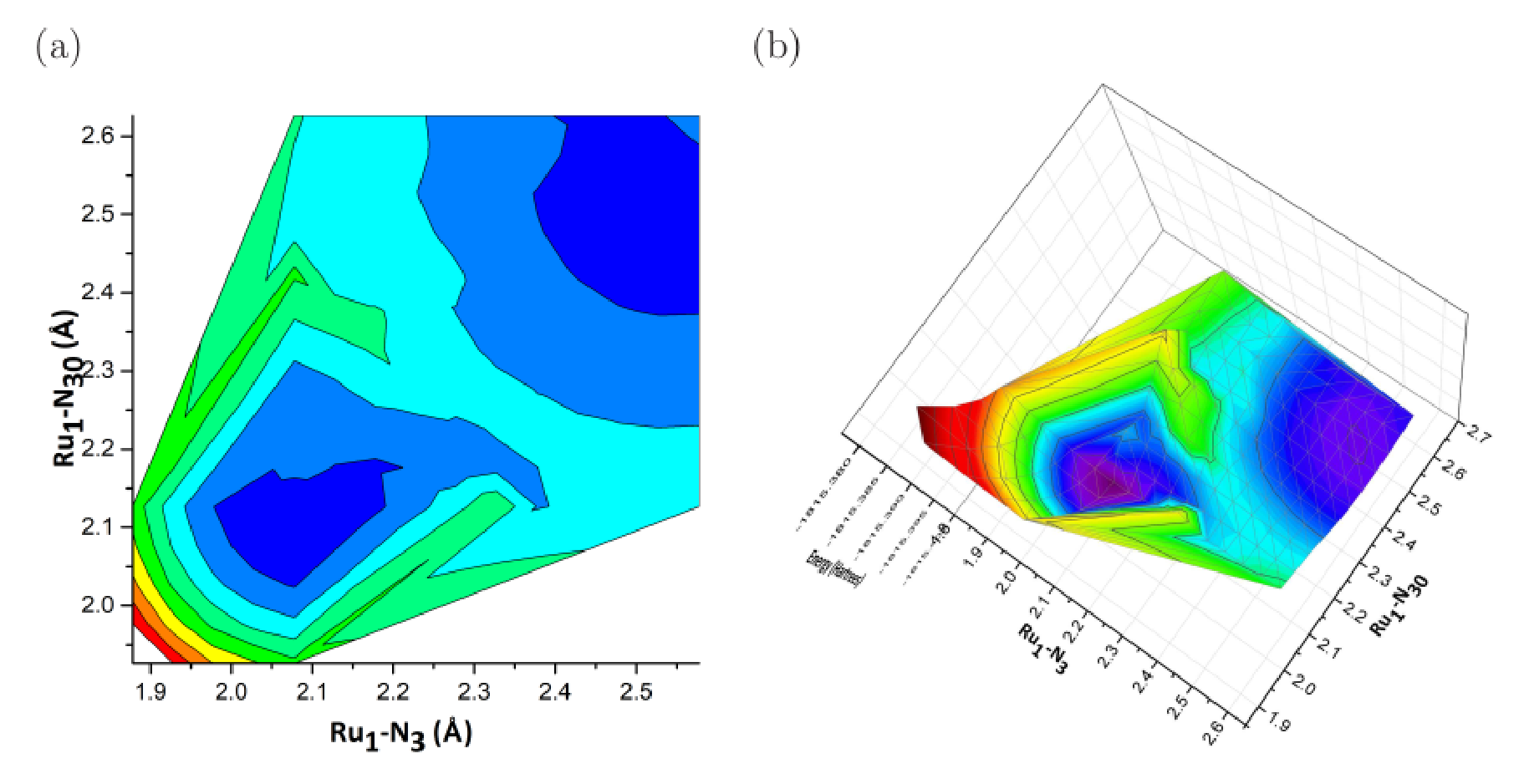} 
\caption{2D scan for complex {\bf 70} using {\sc Gaussian} at the 
B3LYP(VWN3)/6-31G \& LANL2DZ(Ru)/SMD(CH$_3$CN) level: (a) contour plot, (b)
surface. 
\label{fig:70_scan}
}
\end{center}
\end{figure}
NEB and IRC calculations for the {\em trans} $^3$MLCT $\rightarrow$ $^3$MC 
dissociation mechanism require us to optimize the initial $^3$MLCT reactant 
and final $^3$MC product geometries.
Optimization of the $^3$MLCT minimum is relatively straightforward.  It 
suffices to do a vertical excitation from the optimized ground state
geometry and then to allow the molecule to relax on the triplet surface.
This gives a Jahn-Teller distorted geometry with three pairs of bond lengths,
all nearly identical to each other and which we will designate as short.

Finding the $^3$MC minimum for {\em trans} dissociation is more difficult.
We proceed by a 2D scan where each of the two {\em trans} bond lengths are
varied and the geometry is otherwise fully relaxed.  
{\bf Figure~\ref{fig:70_scan}} shows such a scan for complex {\bf 70}.
Similar plots for complexes {\bf 6}, {\bf 73}, and {\bf 74} may be found
in the SI.
Note that the scan should be symmetric upon the reflection through,
respectively, the Ru1-N22 = Ru1-N43 and R1-N3 = Ru-N30 lines, 
but only if the same scan minima are found
on both sides of the line, which is not expected to happen.  However
such scans are good enough to allow us to locate a good guess for 
the $^3$MC {\em trans} minimum which is then optimized to give us
our final $^3$MC geometry.  As these  
scans are computationally intensive, we have only done them 
at the B3LYP(VWN3)/6-31G \& LANL2DZ(Ru)/SMD(CH$_3$CN) level 
for the solvation model.
Indeed, this level is our main workhorse, but we have also carried out 
a few higher-level calculations.  
Bond length information at this level
has been collected in {\bf Table~\ref{tab:geomB3LYP631GLANL2DASMD}}.
\marginpar{\color{blue} s, m, l}
Notice that the optimized $^3$MC has two long (l) bonds {\em trans} to
each other, two short (s) bonds, and two medium (m) length bonds.  
Some {\sc Orca} optimized bond lengths have been given to show the level
of agreement between our {\sc Orca} and {\sc Gaussian} calculations.
\begin{table}
\begin{center}
\begin{tabular}{ccccc}
\hline \hline
Bond & Ground State 
     & $^3$MLCT                  
     & $^3$TS  
     & $^3$MC \\
\hline
\multicolumn{5}{c}{[Ru(bpy)$_3$]$^{2+}$ ({\bf 6})}\\
Ru$_1$-N$_2$  
             & 2.104 (s) 
             & 2.125 (s) [2.122 (s)]
             & 2.113 (s) [2.110 (s)]
             & 2.120 (s) [2.109 (s)] \\
Ru$_1$-N$_3$ 
             & 2.104 (s) 
             & 2.112 (s) [2.102 (s)]
             & 2.113 (s) [2.102 (s)]
             & 2.122 (s) [2.109 (s)] \\
Ru$_1$-N$_{22}$ 
             & 2.104 (s) 
             & 2.078 (s) [2.072 (s)]
             & 2.200 (m) [2.208 (m)]
             & 2.497 (l) [2.445 (l)] \\
Ru$_1$-N$_{23}$ 
             & 2.103 (s) 
             & 2.082 (s) [2.071 (s)]
             & 2.116 (s) [2.115 (s)]
             & 2.182 (m) [2.162 (m)] \\ 
Ru$_1$-N$_{42}$ 
             & 2.103 (s) 
             & 2.113 (s) [2.102 (s)]
             & 2.191 (m) [2.175 (m)]
             & 2.182 (m) [2.162 (m)] \\
Ru$_1$-N$_{43}$ 
             & 2.104 (s) 
             & 2.135 (s) [2.121 (s)]
             & 2.262 (l) [2.258 (l)]
             & 2.503 (l) [2.623 (l)] \\
\multicolumn{5}{c}{[Ru(4,4'-dm-bpy)$_3$]$^{2+}$ ({\bf 70})}\\
Ru$_1$-N$_2$ 
             & 2.103 (s) 
             & 2.126 (s) [2.121 (s)]
             & 2.109 (s) [2.109 (s)]
             & 2.121 (s) [2.108 (s)] \\
Ru$_1$-N$_3$ 
             & 2.104 (s) 
             & 2.078 (s) [2.071 (s)]
             & 2.210 (m) [2.210 (m)]
             & 2.488 (l) [2.450 (l)] \\
Ru$_1$-N$_{4}$ 
             & 2.103 (s) 
             & 2.104 (s) [2.101 (s)]
             & 2.180 (m) [2.180 (m)]
             & 2.175 (m) [2.161 (m)] \\
Ru$_1$-N$_{29}$ 
             & 2.103 (s) 
             & 2.106 (s) [2.100 (s)]
             & 2.100 (s) [2.100 (s)]
             & 2.121 (s) [2.108 (s)] \\ 
Ru$_1$-N$_{30}$ 
             & 2.104 (s) 
             & 2.127 (s) [2.121 (s)]
             & 2.267 (l) [2.266 (l)]
             & 2.519 (l) [2.448 (l)] \\
Ru$_1$-N$_{31}$ 
             & 2.103 (s) 
             & 2.075 (s) [2.071 (s)]
             & 2.116 (s) [2.116 (s)]
             & 2.176 (m) [2.163 (m)] \\
\multicolumn{5}{c}{[Ru(4,4'-dph-bpy)$_3$]$^{2+}$ ({\bf 73})}\\
Ru$_1$-N$_2$ 
             & 2.100 (s) 
             & 2.101 (s) 
             &           
             & 2.489 (l)  \\
Ru$_1$-N$_3$ 
             & 2.101 (s) 
             & 2.126 (s) 
             &           
             & 2.177 (m) \\
Ru$_1$-N$_{4}$  
             & 2.102 (s) 
             & 2.075 (s) 
             &           
             & 2.113 (s) \\
Ru$_1$-N$_{29}$ 
             & 2.100 (s) 
             & 2.102 (s) 
             &           
             & 2.476 (l) \\
Ru$_1$-N$_{39}$ 
             & 2.101 (s) 
             & 2.123 (s) 
             &           
             & 2.175 (m) \\
Ru$_1$-N$_{41}$ 
             & 2.101 (s) 
             & 2.073 (s) 
             &           
             & 2.110 (s) \\
\multicolumn{5}{c}{[Ru(4,4'-DTB-bpy)$_3$]$^{2+}$ ({\bf 74})}\\
Ru$_1$-N$_2$ 
             & 2.104 (s) 
             & 2.106 (s) 
             &           
             & 2.534 (l) \\
Ru$_1$-N$_3$ 
             & 2.102 (s) 
             & 2.125 (s) 
             &           
             & 2.167 (m) \\
Ru$_1$-N$_{4}$ 
             & 2.101 (s) 
             & 2.073 (s) 
             &           
             & 2.117 (s) \\
Ru$_1$-N$_{68}$ 
             & 2.104 (s) 
             & 2.126 (s) 
             &           
             & 2.180 (m) \\
Ru$_1$-N$_{69}$ 
             & 2.102 (s) 
             & 2.077 (s) 
             &           
             & 2.114 (s) \\
Ru$_1$-N$_{70}$ 
             & 2.101 (s) 
             & 2.102 (s) 
             &           
             & 2.476 (l) \\
\hline \hline
\end{tabular}
\caption{
Key Ru-N bond lengths ({\AA}) for different compounds as computed with 
{\sc Gaussian} at the 
B3LYP(VWN3)/6-31G \& Ru(LANLDZ)/SMD(CH$_3$CN) level.
In parentheses: 
``s'' stands for ``short'' ($\sim$ 2.1 {\AA}),
``m'' for ``medium length'' ($\sim$ 2.2 {\AA}), and 
``l'' for ``long'' ($\sim$ 2.3 {\AA} or longer).
A few values computed with {\sc ORCA} are given in square brackets for
comparison purposes.
\label{tab:geomB3LYP631GLANL2DASMD}
}
        \end{center}
\end{table}


For complexes {\bf 73} and {\bf 74}, we also have optimized geometries at
the B3LYP(VWN3)/6-31G(d,p) \& LANL2DZ(Ru)/SMD(CH$_3$CN) 
({\bf Table~\ref{tab:geomB3LYP6311GdpLANL2DASMD}}) level. 
\begin{table}
\begin{center}
\begin{tabular}{ccccc}
\hline \hline
Bond & Ground State 
     & $^3$MLCT                  
     & $^3$TS   
     & $^3$MC \\
\hline
\multicolumn{5}{c}{[Ru(4,4'-dph-bpy)$_3$]$^{2+}$ ({\bf 73})}\\
Ru$_1$-N$_2$ 
             & 2.110 (s) 
             & 2.112 (s) 
             & 
             & 2.508 (l)  \\
Ru$_1$-N$_3$ 
             & 2.111 (s) 
             & 2.142 (s) 
             & 
             & 2.200 (m) \\
Ru$_1$-N$_{4}$  
             & 2.111 (s) 
             & 2.082 (s) 
             & 
             & 2.124 (s) \\
Ru$_1$-N$_{29}$ 
             & 2.110 (s) 
             & 2.112 (s) 
             & 
             & 2.508 (l) \\ 
Ru$_1$-N$_{39}$ 
             & 2.110 (s) 
             & 2.138 (s) 
             & 
             & 2.198 (m) \\
Ru$_1$-N$_{41}$ 
             & 2.111 (s) 
             & 2.081 (s) 
             & 
             & 2.124 (s) \\
\multicolumn{5}{c}{[Ru(4,4'-DTB-bpy)$_3$]$^{2+}$ ({\bf 74})}\\
Ru$_1$-N$_2$ 
             & 2.113 (s) 
             & 2.115 (s) 
             & 
             & 2.511 (l) \\
Ru$_1$-N$_3$ 
             & 2.111 (s) 
             & 2.135 (s) 
             & 
             & 2.199 (m)  \\
Ru$_1$-N$_{4}$ 
             & 2.109 (s) 
             & 2.079 (s) 
             & 
             & 2.124 (s) \\
Ru$_1$-N$_{68}$ 
             & 2.113 (s) 
             & 2.139 (s) 
             & 
             & 2.198 (m) \\ 
Ru$_1$-N$_{69}$ 
             & 2.111 (s) 
             & 2.084 (s) 
             & 
             & 2.125 (s) \\
Ru$_1$-N$_{70}$ 
             & 2.109 (s) 
             & 2.111 (s) 
             & 
             & 2.507 (l) \\
\hline \hline
\end{tabular}
\caption{
Key Ru-N bond lengths ({\AA}) for different compounds as computed with
{\sc Gaussian} at the
B3LYP(VWN3)/6-31G(d,p) \& Ru(LANLDZ)/SMD(CH$_3$CN) level.
In parentheses:
``s'' stands for ``short'' ($\sim$ 2.1 {\AA}),
``m'' for ``medium length'' ($\sim$ 2.2 {\AA}), and
``l'' for ``long'' ($\sim$ 2.3 {\AA} or longer).
\label{tab:geomB3LYP6311GdpLANL2DASMD}
}
        \end{center}
\end{table}



\subsection{$^3$TS and IRC}

TS and IRC calculations are
notoriously difficult for ruthenium(II) polypyridine complexes.  One
reason for this is that there are many different reactions which can
occur and the corresponding TSs are not necessarily very far apart.
An example was reported in Article {\bf III}, namely that 
the reaction pathway was be complicated by the presence of bifurcations.
This is why it is important to have a systematic way to search for
the desired IRCs and TSs.  

We follow the same procedure as previously
described in Article {\bf III}.  Briefly, having found the $^3$MLCT
and $^3$MC minima (step 1), we carry out a NEB calculation with {\sc Orca}
(step 2).  This gives us a first guess as to the $^3$TS in the form of 
the $^3$MEP along the NEB.  This $^3$MEP is only a first approximation to
the $^3$TS.  So {\sc Orca} further optimizes
the $^3$MEP to give a $^3$TS.  At this point, simply because of the way
expertise is distributed among different team members,
calculations are continued with {\sc Gaussian} and the $^3$TS
is reoptimized (step 3).  The nature of the TS is confirmed by calculation
of vibrational frequencies and making sure that there is a single imaginary
vibrational frequency, but this is not enough.  The IRC of the $^3$TS is
further calculated with {\sc Gaussian} to ensure that it leads to the same
$^3$MLCT and $^3$MC geometries as initially used to generate the NEB (step 4).
We also need some way to characterize the pathway.  Naturally, this can be
done in many different ways (we used more than one!), but we have found that
the most useful summary of the pathway is provided by tracing how the six
Ru-N bond lengths vary along the IRC (step 5).  

Calculations are only reported at the basic 
B3LYP(VWN3)/6-31G \& LANL2DZ(Ru)/SMD(CH$_3$CN) level.  Furthermore we were
only able to find the $^3$MEP for complexes {\bf 6}, {\bf 70}, and {\bf 73},
and the $^3$TS was only confirmed via IRC calculation for complexes {\bf 6}
and {\bf 70}.  This is because the SMD implicit solvent calculations are
significantly more compute intensive than are gas-phase calculations.
Furthermore, the mechanism of the {\em trans} $^3$MLCT $\rightarrow$ $^3$MC 
found with the implicit solvent model turns out to be significantly different
than that found in our previous gas-phase calculations (Article {\bf III}).
Nevertheless, the $^3$MEP energy is an upper bound of the $^3$TS energy
and trends in $^3$MEP energies may be expected to be roughly similar to 
trends in $^3$TS energies.

    
\begin{center}
\begin{table}
\footnotesize
\begin{tabular}{ c c c c c }
\hline \hline
Energy{\textbackslash}Complex    & {\bf 6}           & {\bf 70}          & {\bf 73}          & 
                       {\bf 74}          \\
\hline
LI3 (gas)            & 16.78 eV          & 13.78 eV          & 9.68 eV           & 
                       11.97 eV          \\
$^{3}$MLCT           & -1579.52991 Ha    & -1815.39834 Ha    & -2965.61009 Ha    &  -2522.856703
 Ha    \\
    $^{3}$MC             & -1579.52613 Ha    & -1815.39367 Ha    & -2965.60302 Ha     & -2522.851208  Ha    \\
$^{3}$MEP            & -1579.52122 Ha    & -1815.38845 Ha    & -2965.59913 Ha     &
                           \\
$^{3}$TS             & -1579.52238 Ha    & -1815.38962 Ha    &     &
                           \\
$^{3}$MEP-$^{3}$MLCT & 0.00869 Ha        & 0.00989 Ha        & 0.01096 Ha        &
                            \\
                     & 1907.23 cm$^{-1}$ & 2170.60 cm$^{-1}$ & 2405.44 cm$^{-1}$  &
                         \\
$^{3}$TS-$^{3}$MLCT  & 0.00753 Ha        & 0.00872 Ha         &          &
                               \\
                     & 1652.64 cm$^{-1}$  & 1913.82 cm$^{-1}$  &   &
                         \\
$^{3}$MEP-$^{3}$MC   & 0.00491 Ha        & 0.00522 Ha         & 0.00389 Ha        &
                              \\
                     & 1077.62 cm$^{-1}$ & 1145.66 cm$^{-1}$ & 853.76 cm$^{-1}$ &
                        \\
$^{3}$TS-$^{3}$MC    & 0.00374 Ha        & 0.00405 Ha        &         &
                               \\
                     &  823.03 cm$^{-1}$ & 888.87 cm$^{-1}$ &  &
                        \\
$^{3}$MC-$^{3}$MLCT  & 0.00375 Ha       & 0.00467 Ha         & 0.00707 Ha        &
                       0.005495 Ha       \\
                     &  829.61 cm$^{-1}$ & 1024.95 cm$^{-1}$ & 1551.69 cm$^{-1}$  &
                       1206.01 cm$^{-1}$  \\
\hline \hline 
 \end{tabular}
 \caption{Energies for $^{3}$MLCT, $^{3}$MC and TS of all four compounds 
  obtained from B3LYP(VWN3)/6-31G \& LANL2DZ(Ru)/SMD(CH$_3$CN) NEB 
  calculations as implemented in the \textsc{ORCA} code. 
        \label{tab:orcaenergies}}
\end{table}
\end{center}


{\bf Table~\ref{tab:orcaenergies}} 
summarizes key energies and energy differences.
Complex {\bf 6} is the prototypical luminescent ruthenium(II) bipyridine
complex and, as such, has been extensively studied in the literature.
In particular, our value of 1~652.64 cm$^{-1}$ for the 
$^3$MLCT $\rightarrow$ $^3$MC barrier height at the 
B3LYP(VWN3)/6-31G \& LANL2DZ(Ru)/SMD(CH$_3$CN) level
may be compared with the two previously obtained
theoretical values of 1~138 cm$^{-1}$ \cite{SDV+17} 
and 3~040 cm$^{-1}$ \cite{SDAH18} 
also in the same solvent and with the experimental value of
3~800 cm$^{-1}$ \cite{CM83} 
in the same solvent.  This is a reminder that the accuracy of modern
quantum chemical calculations for transition metal complexes is closer to 
5 kcal/mol (1~750 cm$^{-1}$) \cite{FMC+04,FCL+05}
than to the oft quoted 1 kcal/mol goal for chemical accuracy.

Nevertheless, we may hope for consistent trends within a given
theoretical model chemistry (defined by a fixed functional, basis set, etc.)
and hence a qualitativly (even semiquantitatively!) correct description of 
the key {\em trans} $^3$MLCT $\rightarrow$ $^3$MC mechanism. This mechanism
is found to change in the present implicit solvent calculations with respect
to what we reported earlier in gas-phase in Article {\bf III}.
In our gas-phase calculations, to get from the $^3$MLCT minimum to the $^3$MC minimum,
the complex passes through a $^3$TS where both bonds of a single bipyridine ligand are 
dissociating symmetrically.  After the $^3$TS, the IRC moved out upon a
sort of ridge in hyperspace from which it was found to be able to bifurcate to either
of two symmetry-equivalent $^3$MC valleys (i.e., minimum energy geometries).  
One would normally expect to be able to apply the Bell-Evans-Polanyi principle \cite{B36,EP36}
which predicts a linear relationship between the $^3$TS-$^3$MLCT barrier and
the $^3$MLCT-$^3$MC energy difference. In particular, in the double parabola
model described in Article {\bf III}, the height of the $^3$TS-$^3$MLCT barrier 
should decrease as the $^3$MLCT-$^3$MC energy difference increased.
However the Bell-Evans-Polanyi principle is not applicable in the case of the
above mentioned bifurcation.
In fact, we found that the $^3$TS-$^3$MLCT barrier is the same to within
the accuracy of our numerical method.

\begin{figure}
\begin{center}
\includegraphics[width=0.90\textwidth]{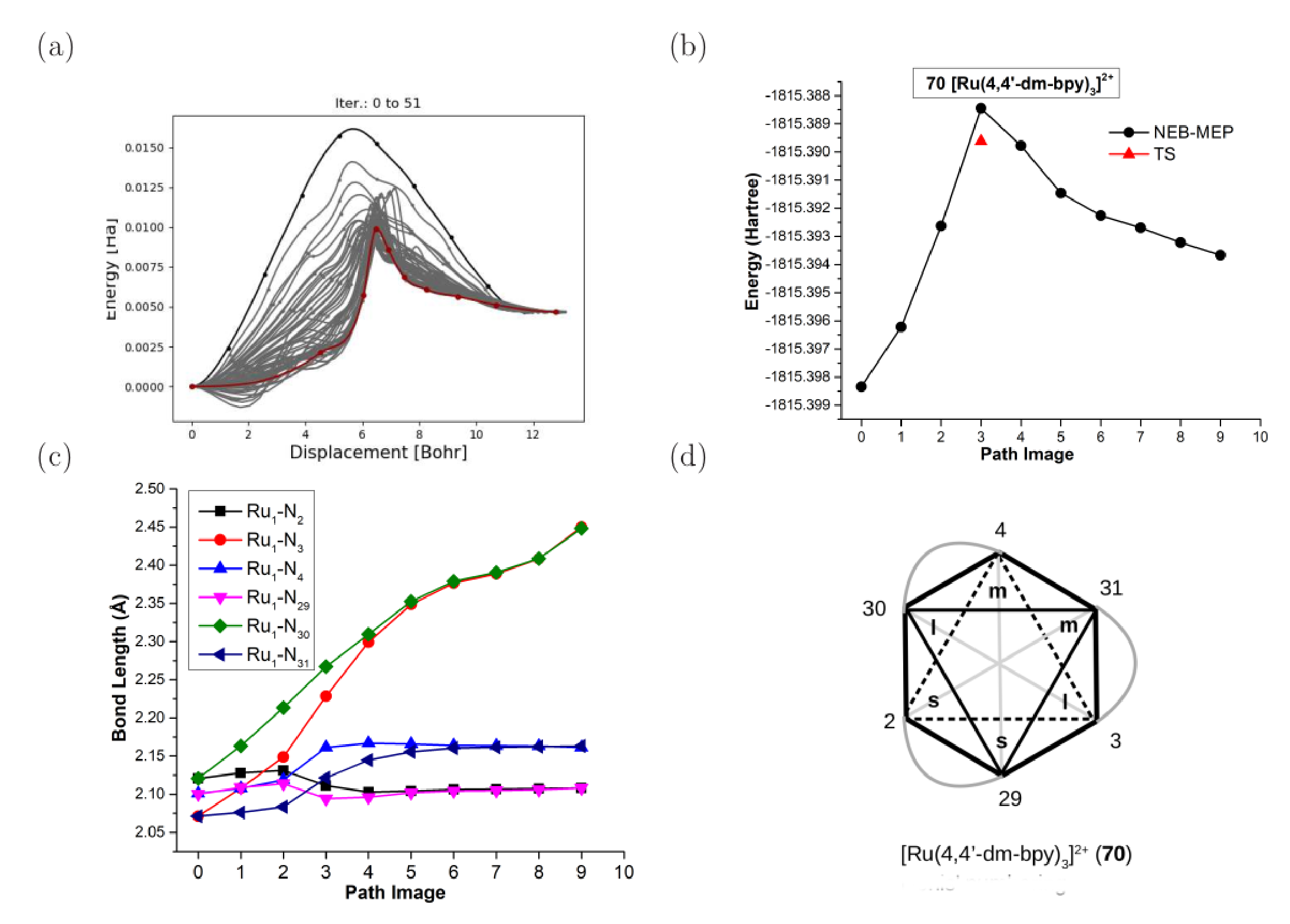} 
\caption{$^3$MLCT $\rightarrow$ $^3$MC NEB scan for complex {\bf 70} at the 
B3LYP(VWN3)/6-31G \& LANL2DZ(Ru)/SMD(CH$_3$CN) 
 level: 
(a) NEB iterations, (b) final NEB, 
(c) corresponding Ru-N bond distances along the NEB,
and (d) cartoon of TS.
\label{fig:70_NEB}
}
\end{center}
\end{figure}

\begin{figure}
\begin{center}
\includegraphics[width=0.90\textwidth]{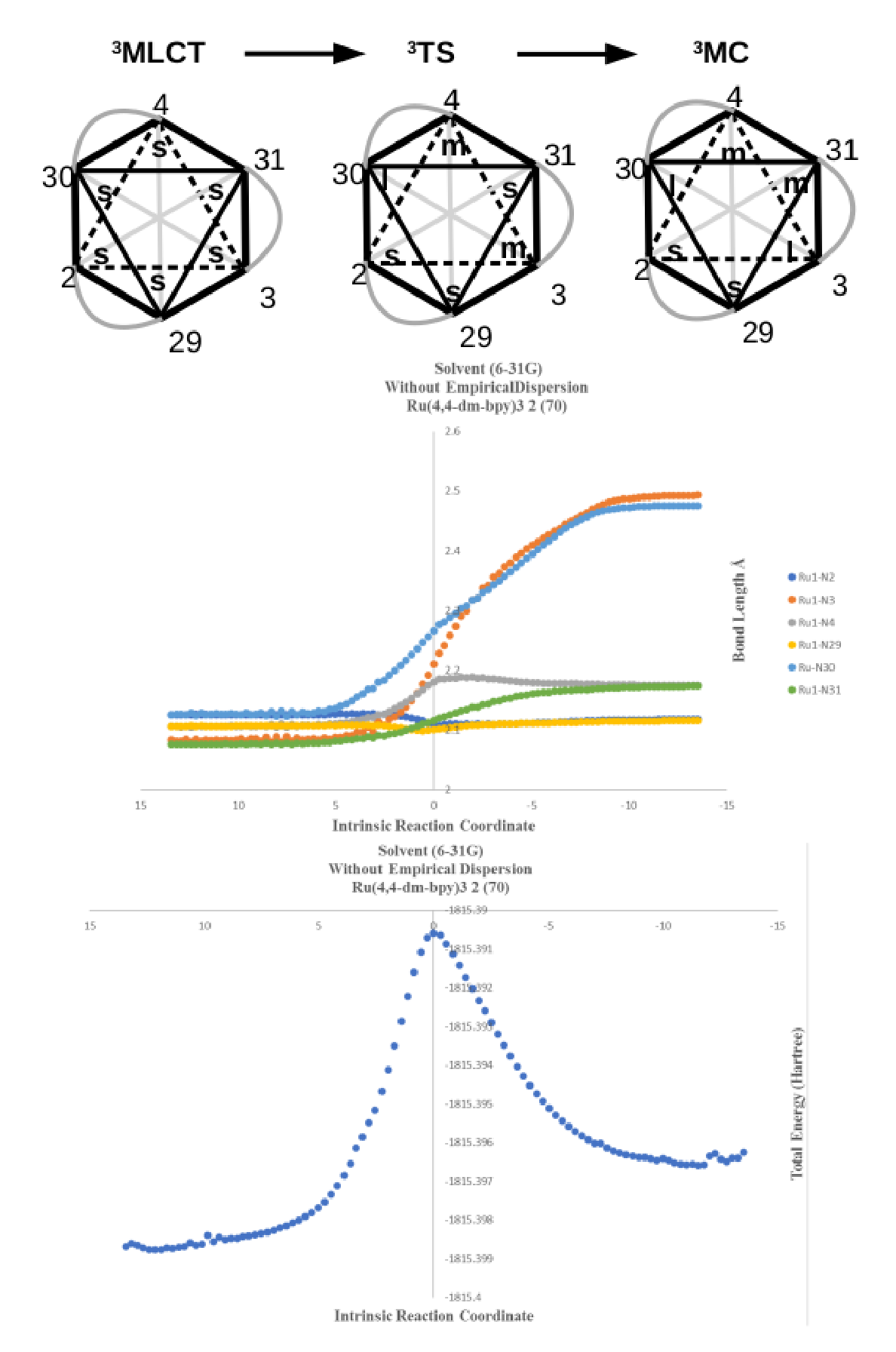} 
\caption{$^3$MLCT $\rightarrow$ $^3$MC IRC scan for complex {\bf 70} at the 
B3LYP(VWN3)/6-31G \&  LANL2DZ(Ru)/SMD(CH$_3$CN) 
level.
\label{fig:70_Aladdin_IRC}
}
\end{center}
\end{figure}

\begin{figure}
\begin{center}
\includegraphics[width=0.90\textwidth]{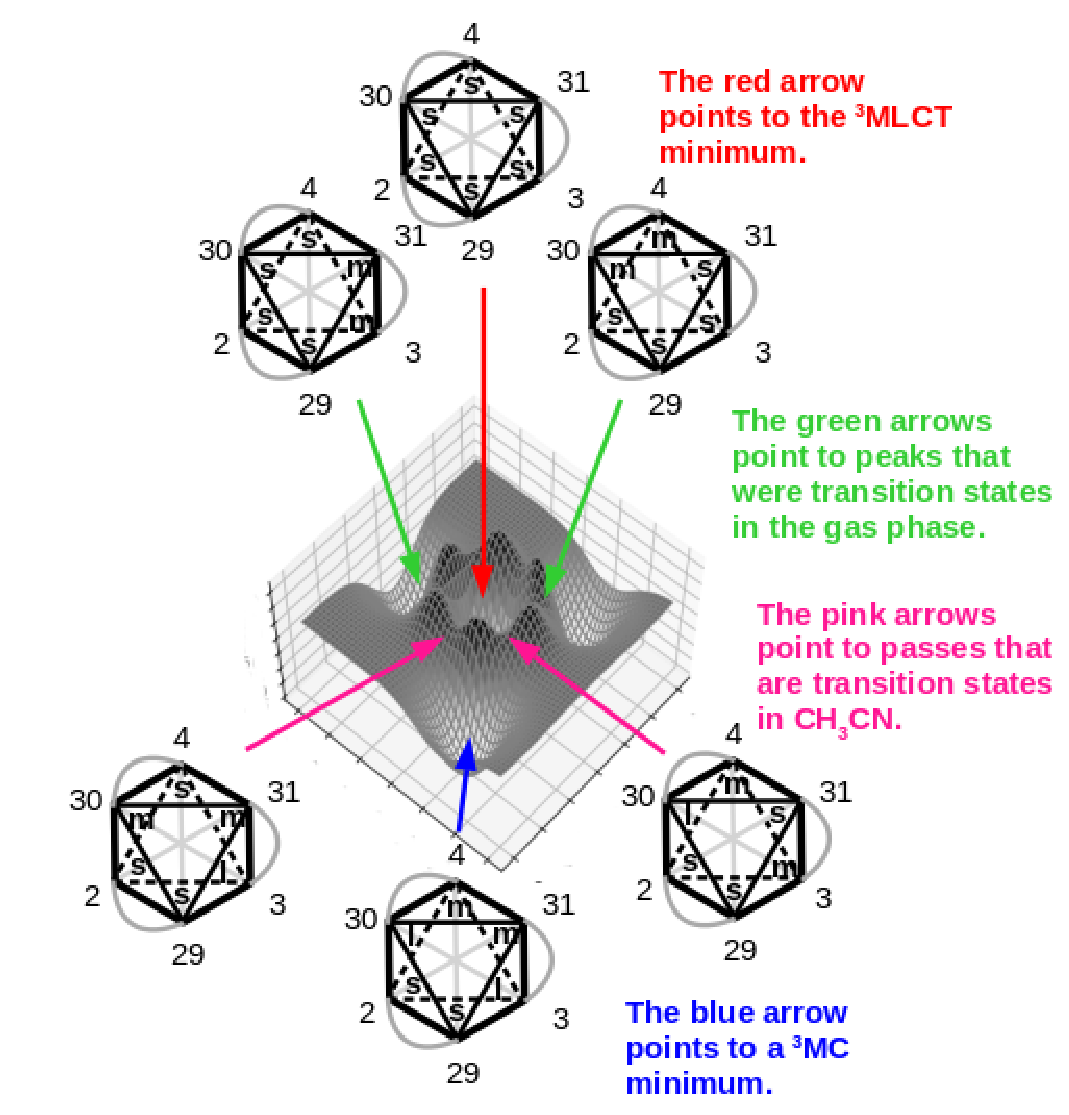} 
\caption{Cartoon showing our understanding of the main features of the triplet
PES for complex {\bf 70}.
\label{fig:PES}
}
\end{center}
\end{figure}

The $^3$MLCT $\rightarrow$ $^3$MC {\em trans}-dissociation reaction in CH$_3$CN
is very different than in the gas phase.  The polar nature of the CH$_3$CN
solvent has two implications.  The first is that the $^3$MLCT minimum is stabilized 
relative to the $^3$MC minimum.  In fact, we find the $^3$MLCT minimum to now have a 
lower energy than the $^3$MC minimum, which is just the opposite of what was found in 
the gas phase.  The second implication of the polar solvent is a symmetry breaking of 
the $^3$TS for the $^3$MLCT $\rightarrow$ $^3$MC {\em trans}-dissociation
reaction.  The $^3$TS is stabilized by lowering its symmetry so that one of the bipyridine 
ligand bonds is now breaking faster than the other for the same ligand.  
These points are illustrated for complex {\bf 70} by the NEB results from {\sc Orca}
shown in {\bf Fig.~\ref{fig:70_NEB}} and the carefully optimized IRC results from 
{\sc Gaussian} shown in {\bf Fig.~\ref{fig:70_Aladdin_IRC}}.  (See the SI for the
corresponding figures for complex {\bf 6}.)  A little thought shows that this 
implies two routes to the same {\em trans}-dissociated product.  
The asymmetry of the TS in our implicit solvent calculation leads us to revise
the cartoon shown in Fig.~6 of Article {\bf III} and to replace it with the new
cartoon shown in {\bf Fig.~\ref{fig:PES}}.
As no bifurcation is observed and we may now expect that the Bell-Evans-Polanyi 
principle to apply.

\subsection{LI3 Orbital-Based Luminescence Index}


We have seen that the original definition of LI3 must be replaced with the new definition
given by LI4 in the solution case, otherwise the values vary too much between gas and
solvated molecules.  But since LI4 in solution is essentially the same as the gas-phase
LI3, we will continue to use our previously calculated values of LI3. 
Let us first review what we have already learned about LI3 from our previous gas-phase
work.  We will then discuss what is the same in our CH$_3$CN solution calculations and 
what is different than in our previous gas-phase calculations.

\begin{figure}
\begin{center}
\includegraphics[width=0.80\textwidth]{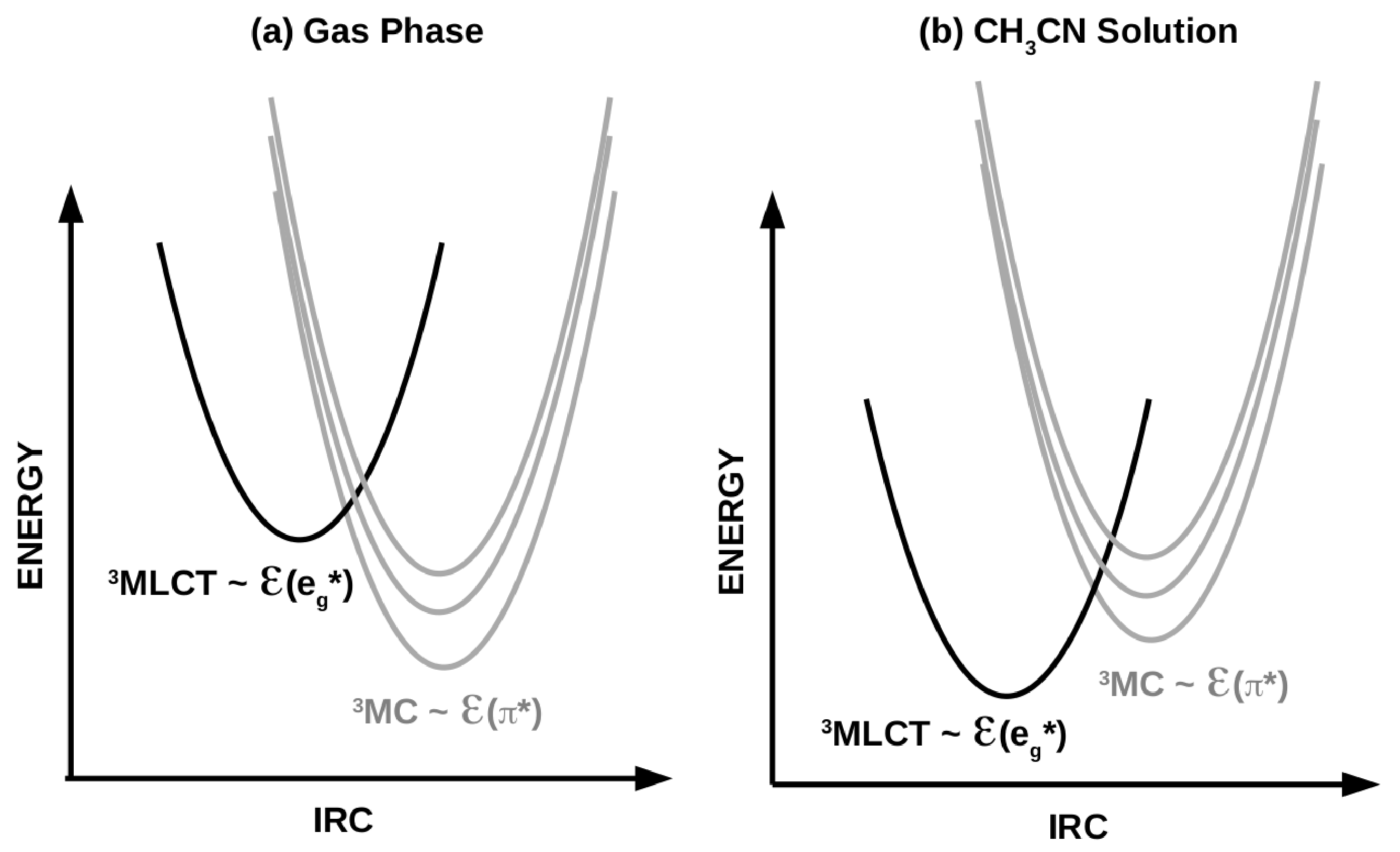}
\caption{Na\"{\i}ve double parabola model.  The position of the $^3$TS
is approximated as where the $^3$MLCT and $^3$MC curves cross.
In either the gas-phase or in CH$_3$CN, increasing the $^3$MLCT-$^3$MC
energy difference (i.e., making it less negative in the CH$_3$CN case)
decreases the $^3$TS-$^3$MLCT energy difference (i.e., the 
$^3$MLCT $\rightarrow$ $^3$MC barrier height).
\label{fig:2parabola} 
} 
\end{center}
\end{figure}
Article {\bf II} proposed the LI3 orbital-based luminescence index as an estimate of the 
$^3$MLCT $\rightarrow$ $^3$MC barrier and a good correlation was found between LI3 and a
$E_{\mbox{ave}}$ ``barrier height'' derived from the temperature-dependence of luminescence
lifetimes of about 100 polypyridine Ru(II) complex condensed-phase measurements.  This is not
a proof that LI3 correlates directly with the real $^3$MLCT $\rightarrow$ $^3$MC barrier height.
Indeed Article {\bf III} investigated the relationship between LI3 and the energetics of the
$^3$MLCT $\rightarrow$ $^3$MC {\em trans}-dissociation reaction in gas phase.  In the gas 
phase,  the $^3$MLCT minimum was found to be at higher energy than that of the 
$^3$MC minimum.  It was found that the $^3$MLCT-$^3$MC energy difference
increased as LI3 increased.  This was explained in Article {\bf III} by the use of
a na\"{\i}ve double parabola model ({\bf Fig.~\ref{fig:2parabola}}) with the energies of 
the products (P) and reactants (R) at the reactant geometry ($x_R$) given by,
\begin{eqnarray}
  E_P(x_R) & \approx & \epsilon_{\pi^*} \nonumber \\
  E_R(x_R) & \approx & \epsilon_{e_g^*} \, ,
  \label{eq:LI.1}
\end{eqnarray}
relative to the energy of the $(t_{2g})^5 (\pi^*)^0 (e_g^*)^0$ configuration.  Then
\begin{equation}
  \mbox{LI3} = \frac{\left[(E_P(x_R)+E_R(x_R))/2 \right]^2 }{E_P(x_R)-E_R(x_R)} \, .
  \label{eq:LI.2}
\end{equation}
[The plus and minus signs were accidently interchanged in Article {\bf III}. See Eq.~(4) of 
that article.]
This shows that increasing LI3 is mainly due to decreasing $E_P(x_R)-E_R(x_R)$ which
translates to decreasing the $^3$MC-$^3$MLCT energy difference or, equivalently, to
increasing the $^3$MLCT-$^3$MC energy difference.

\begin{figure}
\begin{center}
\includegraphics[width=0.80\textwidth]{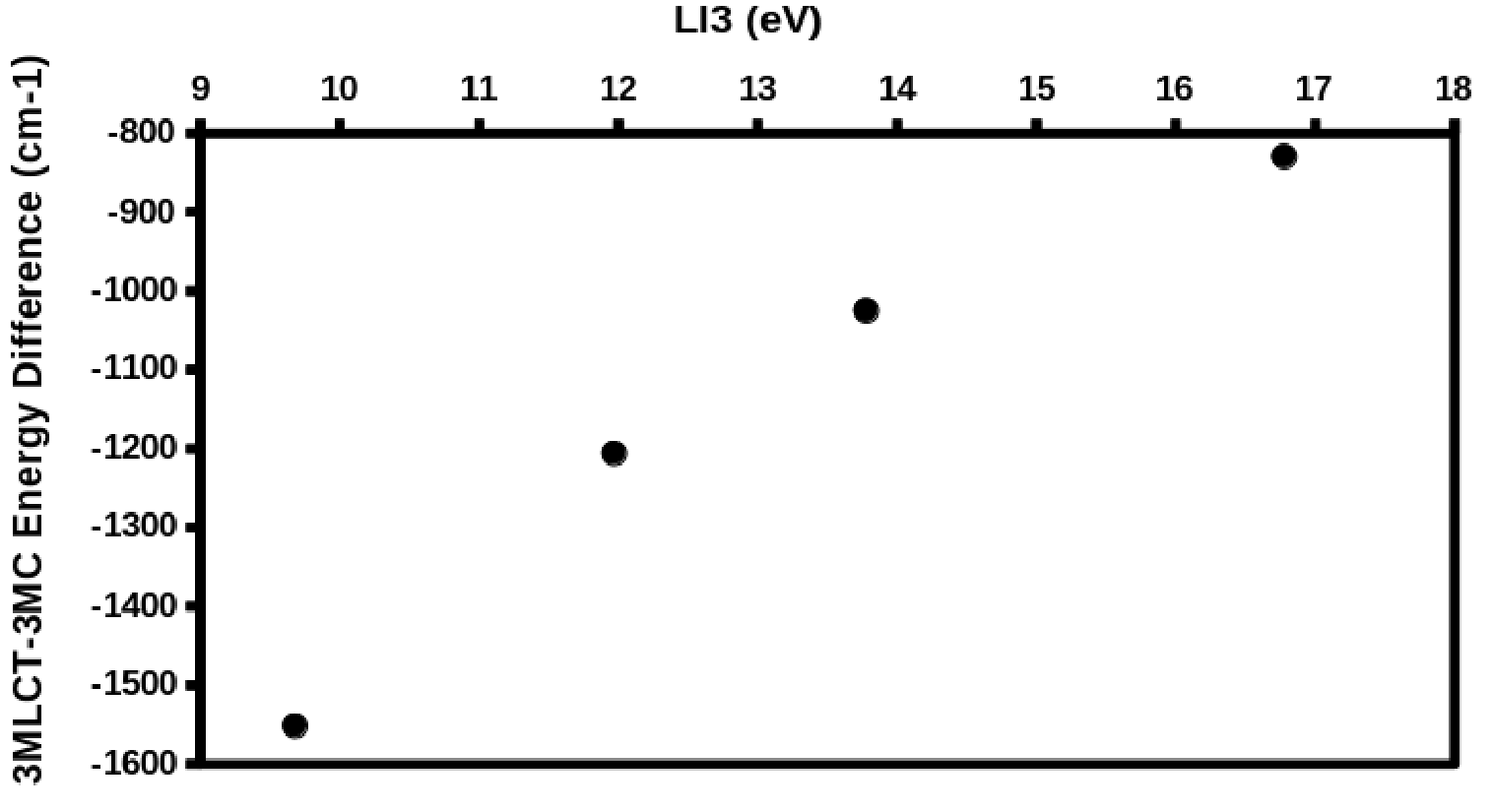}
\caption{Variation of the $^3$MLCT-$^3$MC energy difference calculated at 
the B3LYP(VWN3)/6-31G \& LANL2DZ(Ru)/SMD(CH$_3$CN) 
level as a function of LI3.
\label{fig:minimadifference} 
} 
\end{center}
\end{figure}

\begin{figure}
\begin{center}
\begin{tabular}{cc}
(a) & \\
(b) & \includegraphics[width=0.80\textwidth]{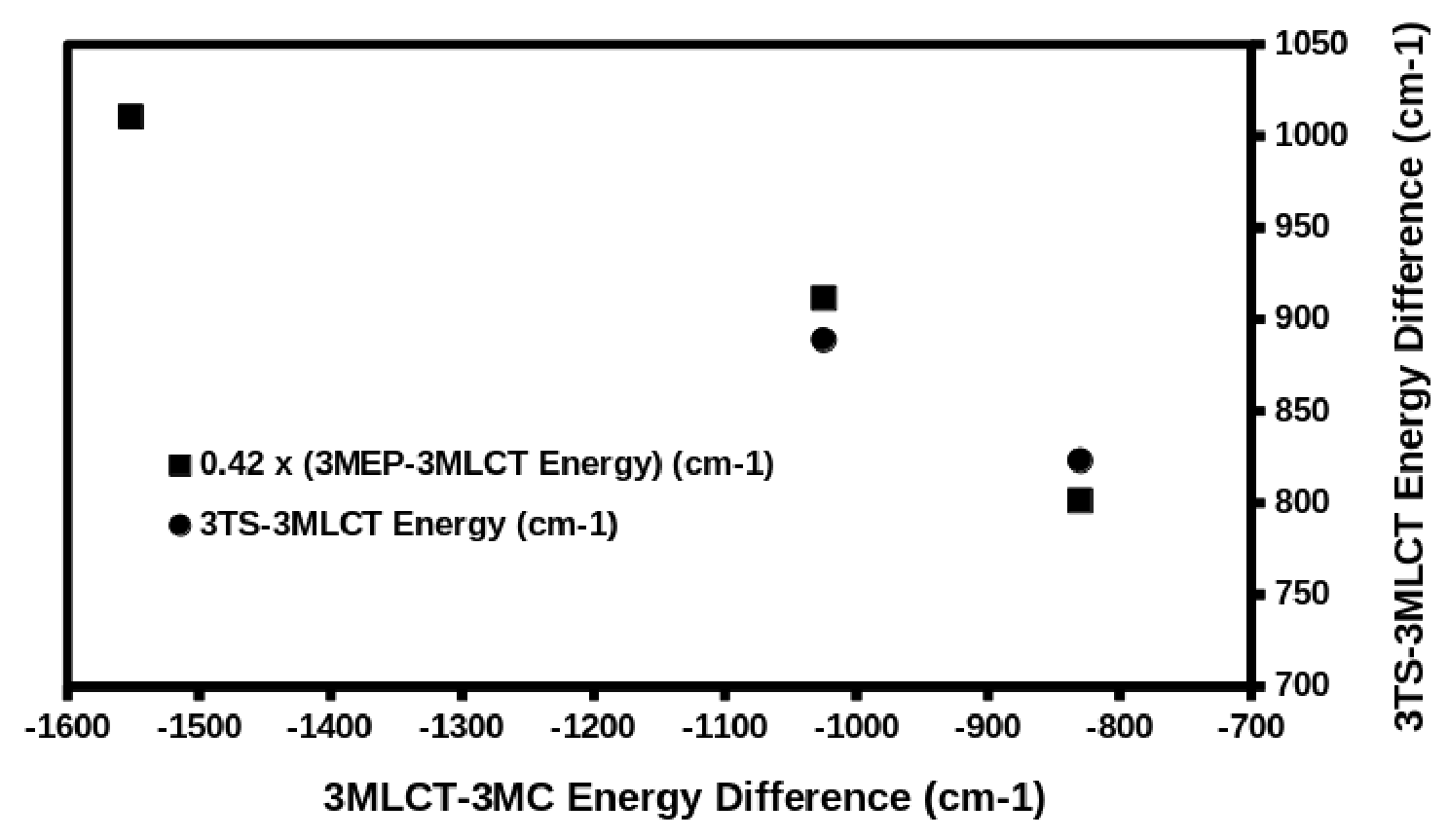} \\
    & \includegraphics[width=0.80\textwidth]{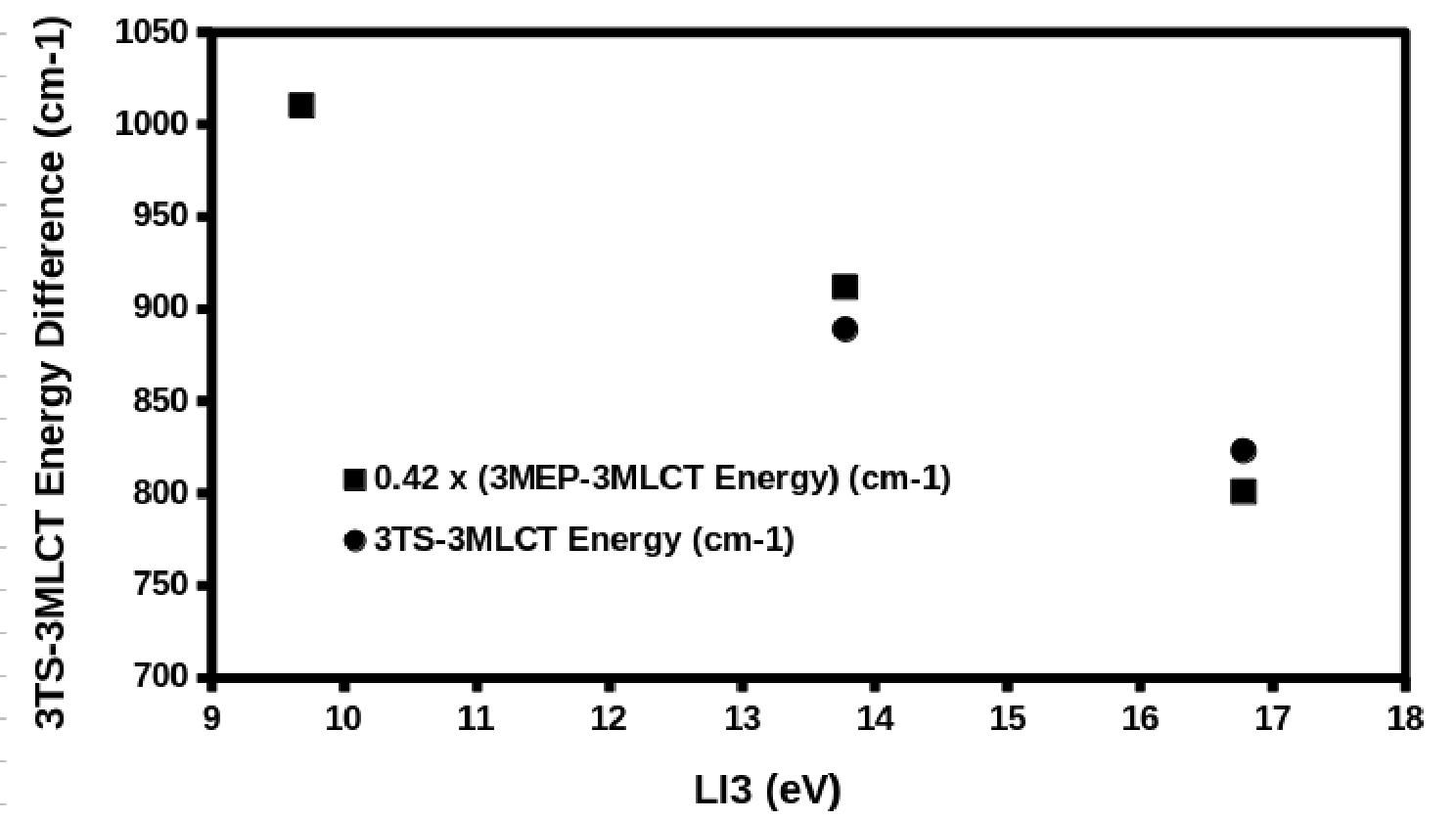} \\
\end{tabular}
\caption{Applicability of the Bell-Evans-Polanyi principle to {\em trans}
dissociation on the triplet B3LYP(VWN3)/6-31G \& 
LANL2DZ(Ru)/SMD(CH$_3$CN) PES. 
Barrier height as a function of: (a) $^3$MLCT-$^3$MC Energy Difference,
(b) LI3.  Here MEP refers to the maximum energy point along the NEB.  This
is typically an overestimate of the TS, so we have taken the liberty of
scaling the $^3$MEP-$^3$MLCT energy difference by a factor of 0.42.
\label{fig:BEP}
} 
\end{center}
\end{figure}

{\bf Figure~\ref{fig:minimadifference}} shows that the $^3$MLCT-$^3$MC energy difference
increases as LI3 increases, just as it does in the gas phase (compare with Fig.~15 of
Article {\bf III}).  The only new thing in CH$_3$CN is that this energy difference is now negative
rather than positive.  As explained above, the absence of a bifurcation on the {\em trans}
$^3$MLCT $\rightarrow$ $^3$MC pathway suggests that the Bell-Evans-Polanyi principle
should apply so that we should see a linear relationship between the $^3$TS-$^3$MLCT
barrier height and both the $^3$MLCT-$^3$MC energy difference and LI3.  As explained
above, the two parabola model predicts that the slope of the graphs should be negative.
This is confirmed in {\bf Fig.~\ref{fig:BEP}} at the 
B3LYP(VWN3)/6-31G \& LANL2DZ(Ru)/SMD(CH$_3$CN) 
level of calculation.  Note that due to
computational resource and time limitations, we have only managed to converge two transition
states.  However we have also included scaled values of the maximum energy points ($^3$MEP)
along the NEB as an estimate of the $^3$TS values.

As discussed above, the orbital-based LI3 luminescence index was 
designed as a qualitative measure of the $^3$TS-$^3$MLCT barrier height.  However the
gas-phase calculations in Article {\bf III} and the present CH$_3$CN implicit solvent
calculations indicate that the $^3$TS-$^3$MLCT barrier height actually decreases 
as LI3 increases.  This is at odds with the traditional explanation of luminescence in
ruthenium(II) polypyridine complexes.  The question is now, ``Why?''

The short answer is that photochemical kinetic models involve many choices and 
assumptions (e.g., Is the mechanism better described by Marcus charge transfer
theory or by Eyring transition state theory and does temperature even make sense in 
this context?)   A longer and very interesting answer has been given by
Hern\'andez-Castillo, Eder, and Gonz\'alez \cite{HEG24}
in a recent review which emphasizes the need to take competing pathways into account 
in order to give a quantitative expanation of luminescence lifetimes in these compounds.  
To some extent, this has been apparent for some time (e.g., see the elaborate model of 
luminescence lifetimes discussed in Article {\bf II} before drastic simplification).
Interestingly Hern\'andez-Castillo, Eder, and Gonz\'alez suggest that the main pathway
for luminescence quenching is by whichever pathway leads most quickly back to the
ground state and that this is {\em not} the traditional {\em trans} dissociation 
mechanism.  At the risk of oversimplification, we suggest that our results are 
consistent with the idea that the {\em trans} dissociation $^3$MLCT $\rightarrow$ $^3$MC 
mechanism may be serving as a reservoir for repopulating the phosphorescent $^3$MLCT 
state and therefore that luminescence lifetimes will actually increase as the 
{\em trans} dissociation $^3$MLCT $\rightarrow$ $^3$MC barrier height decreases.



\section{Conclusion}
\label{sec:conclude}

Mixed quantum/classical (and sometimes completely quantum) photodynamics
modeling represents a state-of-the-art technique for investigating 
photoprocesses such as luminescence in ruthenium(II) polypyridine complexes.
However these more sophisticated techniques suffer not only from their 
heavy use of computational resources, making applications to luminescence
in ruthenium(II) polypyridine complexes extraordinarily rare, but they 
are also far removed from the ligand-field theory (LFT) orbital-based
thinking common among most synthetic chemists.  We have tried to 
remedy this failure in a series of papers ({\bf I} \cite{WJL+14}, 
{\bf II} \cite{MCA+17}, {\bf III} \cite{MDC24}) of which the present
paper is Article {\bf IV}.  In particular, Article {\bf II} showed that
the third orbital-based luminescence index (LI3) that we developed 
based upon the use of frontier-molecular orbital factors that we expected
would govern the $^3$MLCT $\rightarrow$ $^3$MC {\em trans}-dissociation 
barrier height and computed using the partial density of states (PDOS)
method of Article {\bf I} correlated well with experimentally-derived
trends in the $^3$MLCT $\rightarrow$ $^3$MC {\em trans}-dissociation 
barrier height.  However correlation is not causality and so we sought
a deeper explanation of what the LI3 index was actually measuring.
Article {\bf III} studied the gas-phase $^3$MLCT $\rightarrow$ $^3$MC 
{\em trans}-dissociation mechanism.  It was found that the barrier height
was ligand-independent to within the accuracy of our computations but that
LI3 correlates well with the energy difference between the higher-energy
$^3$MLCT state and the lower-energy $^3$MC state.  Although the 
Bell-Evans-Polanyi principle would normally imply that the barrier height
should also correlate with LI3, this was not so because of a novel pathway
involving charge transfer to a single bipyridine ligand which symmetrically
distances itself from the metal atom before then bifurcating to one of two
isomers.  For this same reason, the transition state is particularly
hard to converge.

As most experimental measurements of luminescence lifetimes are done for
complexes in condensed phases, it was felt that it was particularly important
to carry out calculations for the solution phase reaction.  This article
reports these calculations obtained in CH$_3$CN using an implicit solvent
model.  Several things differences between the gas-phase and CH$_3$CN
solution mechanisms.  The solvent stabilizes the $^3$MLCT state so that it
is now {\em lower in energy} than the $^3$MC state.  It also has the effect
of polarizing the $^3$MLCT $\rightarrow$ $^3$MC {\em trans}-dissociation 
transition state ($^3$TS) so that each gas-phase $^3$TS becomes two
$^3$TSs in solution.  This eliminates the bifurcation, making convergence
of the $^3$TS easier except that the solvent model requires additional
computational resources.  The solvent also shifts the PDOS but does not
change its shape, so that we may replace the LI3 luminescence index with
a new LI4 luminescence index which is insensitive to the presence of the
solvent and less sensitive to the choice of functional and basis set.
The LI4 luminescence index also correlates well with both the $^3$MC-$^3$MLCT
energy difference and, consistent with the Bell-Evans-Polanyi prediction,
with the corresponding barrier height.

Surprisingly, the longest luminescence lifetimes are found to be associated
with the {\em lowest} barrier heights, rather than with the highest barrier 
heights as previously believed.  To be fair, photochemical rate theory is
not as well understood as thermal (i.e., ground-state) rate theory.  It is
far from obvious what temperature means for photochemical processes when
applying Eyring transition state theory or whether we are in a recrossing
regime where Marcus charge transfer theory should apply or whether we might
instead be in some intermediate or perhaps even new type of regime.
Hern\'andez, Eder, and Gonz\'alez recently summarized the relevant literature
for the theory of luminescence lifetimes in ruthenium(II) polypyridine
complexes and presented a new, more complete, theory \cite{HEG24}.
At the risk of oversimplfication of their theory, we will summarize it
by saying that luminescence quenching occurs not by reaching the $^3$MC
state but by a rapid return to the singlet ground state by another route.
If we believe that this, then it might mean that a lower {\em trans} barrier
leads to a {\em longer} luminescence lifetime by preventing the molecule
from finding the most efficient route to return to the ground state.
Of course, we must caution that much more investigation needs to be done
before we can be confident in our model for the overall global mechanism
governing luminescence lifetimes, even in the limited class of complexes
constituted by ruthenium(II) polypyridine complexes.


\section*{Supplementary Information}
\label{sec:SI}
Only plots for complex {\bf 70} have been used in the main article.
Among other supplementary information are some plots for other complexes.
\begin{enumerate}
  \item Comparison of {\sc Gaussian} and {\sc Orca} results
  \item PDOS
  \item Contour plots
  \item {\em Trans} dissociation TSs and IRCs
  \item Author contributions
\end{enumerate}

\section*{Acknowledgement}
\label{sec:thanks}

DM and MEC gratefully acknowledge helpful funding from the 
African School on Electronic Structure Methods and Applications (ASESMA),
ASESMANet, and the US-Africa Initiative. 
We are indepted to a number of people for insightful discussions
throughout our studies of orbital-based luminescence indices in
ruthenium(II) polypyridine complexes.  Many of these people have
already been thanked in Article {\bf III} \cite{MDC24} and we, though
we are grateful, we will not repeat the list of names here.  Instead,
we would like to take this opportunity to thank members of the {\sc Orca}
team for helping us to understand what options are needed to obtain the
best possible agreement between {\sc Orca} and {\sc Gaussian} calculations.
In particular, we are indebted to Miquel Garcia~Rates, Frank Neese, 
Christoph Riplinger, Georgi Stoychev, and Frank Wennmohs for helpful
discussions about subtle aspects of numerical algorithms used in the
present article.
We would like to thank Pierre Girard and
S\'ebastien Morin for technical support in the context
of the Grenoble {\em Centre d'Experimentation du Calcul Intensif en Chimie}
({\em CECIC}) computers used for the calculations reported here. 

\section*{Author Information}
\label{sec:authors}

\paragraph{Corresponding Author}
\begin{quote}
\noindent
Ala~Aldin M.\ H.\ M.\ Darghouth$^*$\\
\textit{
College of Sciences, University of Mosul, Al Majmoaa Street, Mosul,
41002 Iraq}\\
E-mail: {\color{blue} aladarghouth@uomosul.edu.iq}
\end{quote}

\paragraph{Authors}
\begin{quote}
\noindent 
Denis Magero\\
\textit{School of Science, Technology and Engineering, 
Department of Chemistry and Biochemistry,
Alupe University, P.O.\ Box 845-50400, Busia, Kenya}\\
E-mail: {\color{blue} dmagero@au.ac.ke}
\end{quote}
\begin{quote}
\noindent Mark E.\ Casida\\
\textit{
Laboratoire de Spectrom\'etrie, Interactions et Chimie th\'eorique (SITh), 
D\'epartement de Chimie Mol\'eculaire (DCM, UMR CNRS/UGA 5250), 
Institut de Chimie Mol\'eculaire de Grenoble (ICMG,
FR2607), Universit\'e Grenoble Alpes (UGA) 301 rue de la Chimie, BP 53, F-38041 Grenoble Cedex
9, France}\\
{\color{blue} orcid.org/0000-0002-9124-0530};\\
E-mail: {\color{blue} mark.casida@univ-grenoble-alpes.fr}
\end{quote}


\bibliographystyle{myaip}
\bibliography{refs}
\end{document}